\documentclass[11pt,a4paper]{article}

\PassOptionsToPackage{hyphens,spaces,obeyspaces}{url}

\usepackage[utf8]{inputenc}
\usepackage[T1]{fontenc}
\usepackage{times}
\usepackage[margin=1in]{geometry}
\usepackage{graphicx}
\usepackage{booktabs}
\usepackage{amsmath,amssymb}
\usepackage{hyperref}
\usepackage{xcolor}
\usepackage{enumitem}
\usepackage{float}
\usepackage[section]{placeins}   
\usepackage{caption}
\usepackage{fancyhdr}
\usepackage{titlesec}
\usepackage{tocloft}
\usepackage[most,listings]{tcolorbox}
\usepackage{listings}
\usepackage{inconsolata}        
\usepackage{url}
\DeclareUrlCommand\code{\urlstyle{tt}}
\usepackage{microtype}
\usepackage{todonotes}

\hypersetup{
    colorlinks=true,
    linkcolor=accentdark,
    citecolor=accentdark,
    urlcolor=accentdark,
}

\lstdefinestyle{pythonstyle}{
    language=Python,
    basicstyle=\ttfamily\footnotesize,
    keywordstyle=\color{accentdark}\bfseries,
    stringstyle=\color{accentmid},
    commentstyle=\color{black!55}\itshape,
    showstringspaces=false,
    breaklines=true,
    breakatwhitespace=false,
    columns=fullflexible,
    keepspaces=true,
    upquote=true
}

\newtcblisting{pythonapi}{
    colback=lightgray,
    colframe=black!20,
    boxrule=0.5pt,
    arc=4pt,
    left=4pt,
    right=4pt,
    top=4pt,
    bottom=4pt,
    listing only,
    listing options={style=pythonstyle}
}

\newcommand{\appendixtocheader}{%
  \addtocontents{toc}{%
    \protect\vspace{20pt}%
    \protect\noindent{\protect\large\protect\bfseries Appendices}%
    \protect\par\protect\vspace{6pt}%
    \protect\hrule\protect\vspace{10pt}%
  }%
  \addtocontents{toc}{\protect\renewcommand{\protect\cftsecfont}{\protect\small\protect\bfseries}}%
  \addtocontents{toc}{\protect\renewcommand{\protect\cftsubsecfont}{\protect\small}}%
  \addtocontents{toc}{\protect\renewcommand{\protect\cftsecpagefont}{\protect\small}}%
  \addtocontents{toc}{\protect\renewcommand{\protect\cftsubsecpagefont}{\protect\small}}%
  \addtocontents{toc}{\protect\setlength{\protect\cftbeforesecskip}{3pt}}%
  \addtocontents{toc}{\protect\setlength{\protect\cftbeforesubsecskip}{1pt}}%
  \addtocontents{toc}{\protect\setcounter{tocdepth}{2}}%
}

\definecolor{accent}{RGB}{254,60,0}            
\definecolor{accentblue}{RGB}{254,60,0}       
\definecolor{accentdark}{RGB}{180,48,0}       
\definecolor{accentmid}{RGB}{210,70,20}       
\definecolor{accentlight}{RGB}{255,240,235}   
\definecolor{lightgray}{RGB}{245,245,248}

\begin{document}

\thispagestyle{empty}

\noindent%
\begin{minipage}[c]{0.24\textwidth}
  \includegraphics[height=1.18cm]{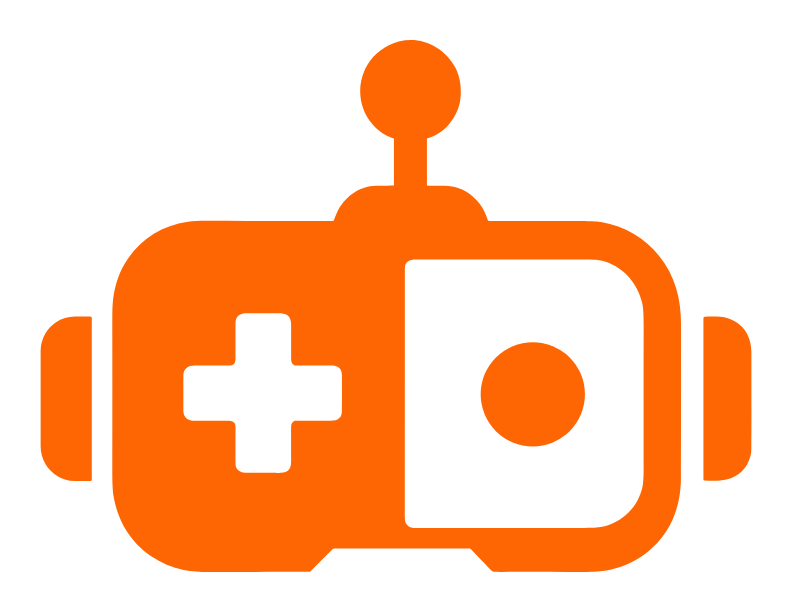}
\end{minipage}%
\hfill%
\begin{minipage}[c]{0.26\textwidth}
  \raggedleft
  {\sffamily\bfseries\fontsize{15}{18}\selectfont\color{accent} GameMind}
\end{minipage}

\vspace{6pt}
{\color{accent}\hrule height 1.2pt}

\vspace{1.4cm}

\begin{center}
    {\fontsize{22}{27}\selectfont\bfseries Cutscene Agent: An LLM Agent Framework for Automated 3D Cutscene Generation}
\end{center}

\vspace{0.8cm}

\begin{center}
    {\large\bfseries Kuaishou GameMind Lab}

    \vspace{0.35cm}

    {\small See \hyperref[sec:contributions]{Contributions} section for a full author list.}
\end{center}

\vspace{0.8cm}

\begin{tcolorbox}[
    enhanced,
    colback=lightgray,
    colframe=lightgray,
    boxrule=0pt,
    borderline west={2.5pt}{0pt}{accentblue},
    arc=0pt,
    outer arc=0pt,
    left=12pt, right=12pt, top=10pt, bottom=10pt,
    fontupper=\normalsize,
]
{\normalsize\bfseries\color{accentblue} Abstract}\par\vspace{6pt}
Cutscenes are carefully choreographed cinematic sequences embedded in video games and interactive media, serving as the primary vehicle for narrative delivery, character development, and emotional engagement. Producing cutscenes is inherently complex: it demands seamless coordination across screenwriting, cinematography, character animation, voice acting, and technical direction, often requiring days to weeks of collaborative effort from multidisciplinary teams to produce minutes of polished content. In this work, we present Cutscene Agent, an LLM agent framework for automated end-to-end cutscene generation. The framework makes three contributions: (1)~a Cutscene Toolkit built on the Model Context Protocol (MCP) that establishes \emph{bidirectional} integration between LLM agents and the game engine---agents not only invoke engine operations but continuously observe real-time scene state, enabling closed-loop generation of editable engine-native cinematic assets; (2)~a multi-agent system where a director agent orchestrates specialist subagents for animation, cinematography, and sound design, augmented by a visual reasoning feedback loop for perception-driven refinement; and (3)~CutsceneBench, a hierarchical evaluation benchmark for cutscene generation. Unlike typical tool-use benchmarks that evaluate short, isolated function calls, cutscene generation requires long-horizon, multi-step orchestration of dozens of interdependent tool invocations with strict ordering constraints---a capability dimension that existing benchmarks do not cover. We evaluate a range of LLMs on CutsceneBench and analyze their performance across this challenging task.
\end{tcolorbox}

\vspace{1cm}

\noindent{\small\color{black!70}\textbf{Project Page:} \url{https://kuaishou-gamemind.github.io/cutscene_agent/}}

\vspace{0.15cm}

\noindent{\small\color{black!70}\textbf{Correspondence:} \href{mailto:ganqi@kuaishou.com}{ganqi@kuaishou.com}}

\newpage

\renewcommand{\contentsname}{Contents}
\tableofcontents

\newpage

\section{Introduction}
\label{sec:introduction}
Automating the creation of high-quality cinematic content has long been a compelling goal for the game and film industries. Cutscenes---carefully choreographed cinematic sequences that deliver narrative, character development, and emotional engagement---represent one of the most resource-intensive aspects of modern game development. Producing a single minute of polished cutscene content typically requires days of collaborative effort from multidisciplinary teams spanning directors, cinematographers, animators, voice actors, and technical artists. This complexity creates substantial barriers for smaller studios and limits iteration speed even for well-resourced productions.

Recent advances in generative AI have opened multiple avenues toward automating cinematic content creation. Diffusion-based text-to-video systems such as Kling~\cite{kuaishou2024kling} can now generate high-resolution cinematic footage from text prompts, but their outputs are raw pixel sequences with no underlying scene structure. In parallel, the emergence of Large Language Models (LLMs) with sophisticated reasoning and planning capabilities has inspired LLM-based orchestration of virtual film production. FilmAgent~\cite{xu2025filmagent} proposes a multi-agent collaborative framework where LLM agents assume film crew roles---director, screenwriter, actor, cinematographer---to generate virtual films through iterative dialogue refinement. MovieAgent~\cite{movieagent2025} extends this paradigm with multi-agent Chain-of-Thought planning for multi-scene, multi-shot long-form video generation. However, these approaches share a fundamental limitation: their outputs are either structured JSON files or pre-rendered videos rather than editable production assets. We term this the editability gap---generated content cannot be modified without complete re-generation, exists outside the game engine pipeline, and discards the underlying semantic structure (keyframes, animation curves, timing data) that enables meaningful professional editing.

In this work, we introduce \textbf{Cutscene Agent}, an LLM agent-driven framework that bridges this gap by generating cutscenes directly as native game engine assets. Our key insight is that effective agent-engine integration must be bidirectional: agents should not only invoke operations but maintain real-time awareness of scene state to make coherent sequential decisions. We achieve this through the Model Context Protocol (MCP), with a dedicated MCP server embedded within Unreal Engine providing real-time access to character management, camera control, and sequence manipulation APIs. Supplementary external services---including speech synthesis and generative AI asset creation---are integrated as additional MCP servers under the same protocol interface. Before each agent decision, the system automatically queries and injects the current Level Sequence state into the agent's context, enabling coherent multi-step generation. Importantly, the exposed MCP tool signatures carry no engine-specific assumptions: porting the agent stack to an alternative backend---such as Unity or Blender---requires only a compatible toolset implementation on the engine side, with no modifications to the agent logic.

\begin{figure}[htbp]
    \centering
    \includegraphics[width=\textwidth]{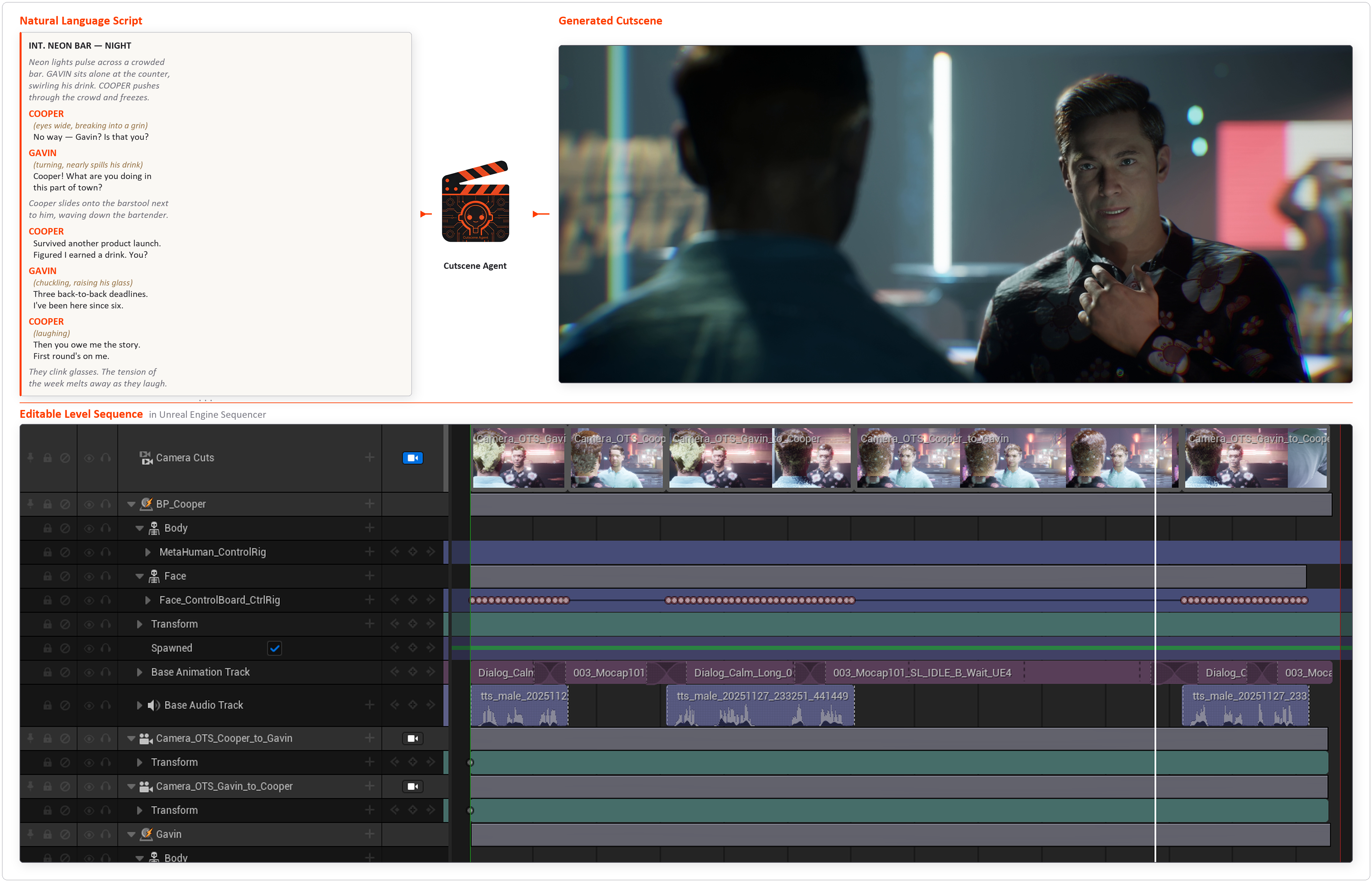}
    \caption{Cutscene Agent generates editable 3D cutscenes from natural-language scripts in minutes, producing multi-track Level Sequences with coordinated character animation, dialogue, and cinematography that remain fully editable by artists. Characters in the demo are taken from MetaHuman assets. See our project page for video results and more examples.}
    \label{fig:teaser}
\end{figure}

Our framework outputs native Unreal Engine Level Sequences---the industry-standard format for UE cinematic content---that can be directly previewed, edited using standard Sequencer tools, extended with additional effects, and exported for various platforms. This engine-native, editable output design fundamentally distinguishes our approach from prior work that treats AI-generated content as final rendered output rather than editable intermediate assets. Figure~\ref{fig:teaser} illustrates the end-to-end pipeline.

We make the following contributions:

\begin{itemize}
    \item \textbf{Cutscene Toolkit.} We design and implement a comprehensive MCP-based interface library that exposes the full cutscene authoring pipeline---character management, asset querying, parameterized camera templates, and real-time scene perception---as composable tools. The toolkit is deliberately engine-agnostic.

    \item \textbf{Agent System.} Built upon the toolkit, we implement an agent system that addresses the dual challenges of long-horizon context management and creative workflow orchestration. A composable, priority-driven prompt manager with automatic state injection and category-aware history compression sustains coherent reasoning over sessions of 60+ sequential tool calls. A hierarchical subagent framework---comprising preset specialist subagents (animation, cinematography, sound design) and dynamically constructed custom subagents---decomposes the monolithic generation task into isolated, domain-focused subtasks. A closed-loop visual reasoning mechanism enables vision-capable subagents to perceive rendered frames and iteratively refine camera composition and character staging through a perceive--reason--act cycle.

    \item \textbf{CutsceneBench.} Cutscene generation requires long-horizon orchestration of dozens of interdependent tool invocations with strict ordering constraints---a complexity dimension that existing tool-use benchmarks, which typically evaluate short, isolated function calls, do not cover. We introduce CutsceneBench, a three-layer hierarchical evaluation framework that assesses tool-call correctness, sequence structural integrity, and narrative quality. We evaluate eight LLMs across 65 scenarios spanning five complexity tiers, revealing substantial variation in multi-step orchestration capability and providing a challenging new benchmark for agentic LLM evaluation.
\end{itemize}

The remainder of this paper is organized as follows. Section~\ref{sec:related} reviews related work in virtual production, AI-driven content generation, tool-augmented agents, and agent evaluation benchmarks. Section~\ref{sec:architecture} presents our system architecture, including the MCP-based toolkit, the agent system, and key technical challenges. Section~\ref{sec:evaluation} introduces our hierarchical evaluation framework and reports benchmark results across multiple LLMs.


\section{Related Work}
\label{sec:related}

Our work builds upon advances in four interconnected research areas: virtual film production and computational cinematography, AI-driven content generation for video and animation, agent-environment interaction and tool-use protocols, and evaluation benchmarks for LLM agents. We review each in turn, highlighting how Cutscene Agent addresses limitations of prior approaches.

\subsection{Virtual Film Production and Computational Cinematography}

The automation of cinematographic processes has been a long-standing goal in computer graphics research. Christie et al.~\cite{christie2008camera} provided an influential survey classifying camera control approaches---from interactive manipulation to fully declarative constraint-based methods---and identifying core challenges including occlusion management and communicative goal modeling. A significant conceptual advancement came with the Toric coordinate system~\cite{lino2015toric}, which provides a compact, principled representation for describing camera compositions relative to two subjects in 3D space, enabling efficient viewpoint computation and smooth interpolation. This representation has been widely adopted in subsequent cinematography research.

Deep learning has substantially advanced automated camera control. Jiang et al.~\cite{jiang2020example} combined Toric representations with Mixture-of-Experts models to extract camera behaviors from real film clips and apply them to 3D animations. Their follow-up work~\cite{jiang2021camera} introduced neural keyframe-based control, using gating LSTMs to encode camera styles while satisfying user-specified keyframes. Pueyo et al.~\cite{pueyo2024cinempc} applied Model Predictive Control to autonomous UAV cinematography, innovatively controlling both camera extrinsics (position, orientation) and intrinsics (focal length, aperture, focus distance)---a capability previously unavailable in automated systems. More recently, Jiang et al.~\cite{jiang2024camera} introduced a transformer-based diffusion model for generating diverse camera trajectories conditioned on textual descriptions, representing an evolution toward text-driven cinematography.

The most recent works have begun exploring language-model-driven camera control. ChatCam~\cite{chatcam2024} introduces CineGPT, a GPT-based autoregressive model for text-conditioned camera trajectory generation rendered on neural radiance field representations. GenDoP~\cite{zhang2025gendop} trains an autoregressive Transformer on DataDoP---a large-scale dataset of 29K real-world shots with camera trajectories, depth maps, and directorial intent captions. These emerging approaches demonstrate growing interest in bridging natural language and cinematographic execution, though they operate on neural scene representations rather than production game engine environments.

Despite these advances, existing work has focused on isolated aspects of cinematography---primarily camera trajectory generation. However, a complete cutscene is far more than camera work: it requires the coordinated orchestration of character spawning, spatial blocking, skeletal animation, dialogue audio, facial performance, camera composition, and cut timing into a single, coherent, editable asset within a game engine. To the best of our knowledge, no prior work addresses the fully automated generation of such complete cutscene assets suitable for direct use in game or film production. Our work aims to fill this gap.

\subsection{AI-Driven Content Generation for Video and Animation}

Diffusion-based text-to-video (T2V) generation has progressed from early explorations---extending pre-trained image diffusion models with temporal layers~\cite{singer2022makeavideo,ho2022imagenvideo,blattmann2023align}---to industrial-scale systems built on Diffusion Transformer (DiT) architectures. Proprietary models such as Sora~\cite{openai2024sora} and Kling~\cite{kuaishou2024kling} now generate high-resolution video with complex motion, synchronized audio, and instruction-based editing. Open-source efforts including HunyuanVideo~\cite{kong2024hunyuanvideo}, Wan~\cite{wan2025}, and Open-Sora~\cite{zheng2024opensora} have rapidly closed this gap, achieving comparable quality at a fraction of the training cost.

A parallel line of research leverages LLM agents to orchestrate film production. FilmAgent~\cite{xu2025filmagent} represents the most directly related work, proposing a multi-agent collaborative framework where LLM agents assume film crew roles---director, screenwriter, actor, and cinematographer---and collaborate through Critique-Correct-Verify and Debate-Judge mechanisms in a Unity 3D sandbox. VideoDirectorGPT~\cite{lin2023videodirectorgpt} uses GPT-4 as a video planner to decompose narratives into structured scene descriptions with entity bounding boxes and consistency groupings for a Layout2Vid generation module. Anim-Director~\cite{li2024animdirector} employs large multimodal models as autonomous agents for controllable animation video generation through script generation, image creation, and video production stages. Most recently, MovieAgent~\cite{movieagent2025} introduces multi-agent Chain-of-Thought planning for multi-scene, multi-shot long-form video generation with character consistency. Complementary research on individual cutscene elements---such as text-to-motion synthesis~\cite{tevet2023mdm,jiang2023motiongpt,qing2023story} and audio-driven facial animation---has also advanced rapidly, but these methods produce raw data requiring retargeting and engine import rather than production-ready assets.

Despite this progress, both paradigms share a fundamental limitation: their outputs are either structured JSON files or pre-rendered pixels rather than editable production assets. Generated content cannot be modified without complete re-generation, exists outside game engine pipelines, and discards the underlying semantic structure (keyframes, animation curves, timing data) that enables meaningful professional editing. Our work addresses this gap by generating native Unreal Engine Level Sequences---editable, extensible assets within a production engine.

\subsection{Agent-Environment Interaction and Tool-Use Protocols}

The integration of external tools with LLM agents has evolved from basic prompting paradigms to sophisticated communication protocols. Chain-of-thought prompting~\cite{wei2022chain} established that providing intermediate reasoning steps unlocks complex reasoning in LLMs, laying the foundation for agent-based planning. Toolformer~\cite{schick2023toolformer} demonstrated that language models can learn to autonomously decide when and how to call external APIs through self-supervised training, achieving competitive performance with much larger models. The ReAct framework~\cite{yao2023react} introduced a paradigm that interleaves reasoning traces with action execution, enabling agents to dynamically plan and adjust based on environmental feedback.

A key paradigm shift came with LLMs as orchestrators of complex tool ecosystems. HuggingGPT~\cite{shen2023hugginggpt} demonstrated ChatGPT orchestrating HuggingFace models through a four-stage pipeline (task planning $\rightarrow$ model selection $\rightarrow$ execution $\rightarrow$ response generation). ViperGPT~\cite{suris2023vipergpt} generates Python programs that compose vision modules for visual reasoning---establishing the code-as-tool paradigm. Gorilla~\cite{patil2023gorilla} and ToolLLM~\cite{qin2024toolllm} addressed scaling tool use to massive API ecosystems (16,000+ APIs), with retrieval-aware training and depth-first search reasoning respectively. In embodied settings, Voyager~\cite{wang2023voyager} demonstrated lifelong learning in Minecraft through an auto-curriculum and ever-growing skill library, while Reflexion~\cite{shinn2023reflexion} introduced verbal self-reflection as a ``semantic gradient'' for iterative agent improvement.

The widespread adoption of structured function calling~\cite{openai2023function} has established a standard mechanism for LLMs to invoke external tools via validated JSON outputs. Building on this, modern agentic frameworks provide higher-level abstractions for multi-step planning, subagent delegation, and conversation management, making it practical to build complex agent systems that chain dozens of tool calls over long horizons. In parallel, the concept of \textit{skills}---reusable, self-contained capability modules that encapsulate domain knowledge, prompt templates, and tool configurations---has emerged as a key pattern for scaling agent capabilities. On the protocol layer, the Model Context Protocol (MCP)~\cite{anthropic2024mcp} provides a standardized interface for exposing tool definitions and resources to LLM agents, decoupling tool implementation from agent logic and enabling modular, server-based tool ecosystems. Our framework adopts MCP as the communication layer between the agent and Unreal Engine, exposing character management, camera control, and sequence manipulation as MCP tools hosted on an in-engine server. Notably, we also provide a set of queryable tools for the agent to interact with the current Level Sequence state, ensuring bidirectional communication.

\subsection{Evaluation Benchmarks for LLM Agents}

Evaluating LLM-based agents has become an active area of research, with benchmarks spanning function calling, multi-step task automation, and open-ended environments.

\textbf{Function calling benchmarks} focus on single-step tool invocation correctness. API-Bank~\cite{li2023apibank} provides 73 runnable APIs with annotated dialogues, evaluating planning, retrieval, and execution. Gorilla~\cite{patil2023gorilla} introduced APIBench with AST-based evaluation for detecting API call hallucination. The Berkeley Function Calling Leaderboard (BFCL)~\cite{patil2025bfcl} has become the de facto standard, using AST-based evaluation across single-turn, parallel, multi-turn, and agentic scenarios---revealing that while models excel at single-turn calls, memory and long-horizon reasoning remain open challenges.

\textbf{Multi-step task benchmarks} evaluate tool composition and dependency handling. TaskBench~\cite{shen2024taskbench} formalizes inter-tool dependencies as a \textit{Tool Graph} (nodes = tools, edges = dependencies) and evaluates task decomposition, tool invocation, and parameter prediction using a Back-Instruct dataset construction method. ToolSandbox~\cite{lu2024toolsandbox} introduces \textit{stateful} tool execution with implicit state dependencies and a built-in user simulator, revealing that complex state-dependent scenarios challenge even state-of-the-art models.

\textbf{End-to-end agent benchmarks} assess performance in realistic, long-horizon environments. AgentBench~\cite{liu2023agentbench} evaluates 29 LLMs across 8 diverse environments, revealing significant gaps between commercial and open-source models. $\tau$-bench~\cite{sierra2024tau} tests agents in stateful retail and airline domains with simulated human interaction, finding that GPT-4 achieves fewer than 50\% success and drops to $\sim$25\% across 8 repeated attempts. WebArena~\cite{zhou2024webarena} provides self-hostable web environments where the best GPT-4 agent achieves only 10.59\% task success versus human performance of 78.24\%. SWE-bench~\cite{jimenez2024swebench} tests agents on 2,294 real GitHub issues requiring multi-file code edits.

\textbf{LLM-as-Judge} approaches address evaluation of subjective quality. Zheng et al.~\cite{zheng2023judging} established this paradigm with MT-Bench and Chatbot Arena, demonstrating that GPT-4 achieves over 80\% agreement with human judgments---matching inter-human agreement levels. G-Eval~\cite{liu2023geval} combines chain-of-thought reasoning with form-filling evaluation, achieving substantially higher correlation with human judgments than traditional metrics.

Our evaluation framework, CutsceneBench (Section~\ref{sec:evaluation}), addresses gaps that none of these benchmarks cover: (1) the absence of canonical ground truth for creative tasks admitting multiple valid solutions, (2) long-horizon trajectories of 10--60 tool calls with cumulative stateful side effects, (3) multi-dimensional quality spanning technical correctness through creative quality, and (4) domain-specific evaluation requiring understanding of cinematographic conventions and game engine semantics.

\section{System Architecture}
\label{sec:architecture}

This section presents the architectural design of Cutscene Agent. We begin with the \textbf{Cutscene Toolkit}---a comprehensive MCP-based interface library for Unreal Engine that enables bidirectional communication between AI agents and the game engine. We then describe the \textbf{Agent System} built upon this toolkit, detailing prompt and context management, subagent delegation, and visual reasoning feedback mechanisms. Finally, we discuss \textbf{key technical challenges} encountered in building a robust in-engine generation system and our approaches to addressing them.

\subsection{Cutscene Toolkit}
\label{sec:toolkit}

A core contribution of this work is the \textbf{Cutscene Toolkit}, a comprehensive interface library that bridges LLM agents and Unreal Engine's cinematic authoring pipeline. Unlike previous approaches relying on predefined action libraries or offline rendering pipelines, our toolkit implements the Model Context Protocol (MCP)~\cite{anthropic2024mcp} to enable real-time, bidirectional communication with UE's Level Sequence system. Notably, the Cutscene Toolkit represents a standalone contribution with broad applicability---it can benefit any Unreal Engine developer seeking to integrate LLM-based automation, independent of our agent system.

The toolkit is organized into four functional modules, each addressing a distinct aspect of cutscene authoring:
\begin{enumerate}[nosep]
    \item \textbf{Character \& Track Management} (\S\ref{sec:toolkit_character})---managing the full lifecycle of characters in a Level Sequence, from spawning actor bindings to populating animation, audio, and facial expression tracks.
    \item \textbf{Asset Management \& Query} (\S\ref{sec:toolkit_asset})---a dual-layer system for static asset storage and dynamic asset ingestion (runtime-generated), with a unified query interface that enforces public/private data separation.
    \item \textbf{Camera Management} (\S\ref{sec:toolkit_camera})---camera creation and a parameterized template system that translates cinematic shot language into precise camera poses computed from character skeletal data.
    \item \textbf{Scene Perception \& Interaction} (\S\ref{sec:toolkit_perception})---tools for serializing sequence state, managing semantic metadata, and interacting with the editor viewport for visual verification.
\end{enumerate}
Additionally, the toolkit integrates with external MCP services (e.g., text-to-speech, audio-driven facial synthesis) that generate dynamic assets consumed through the asset management layer, enabling a modular multi-server architecture where each service contributes specialized capabilities.

\begin{figure}[htbp]
    \centering
    \includegraphics[width=\textwidth]{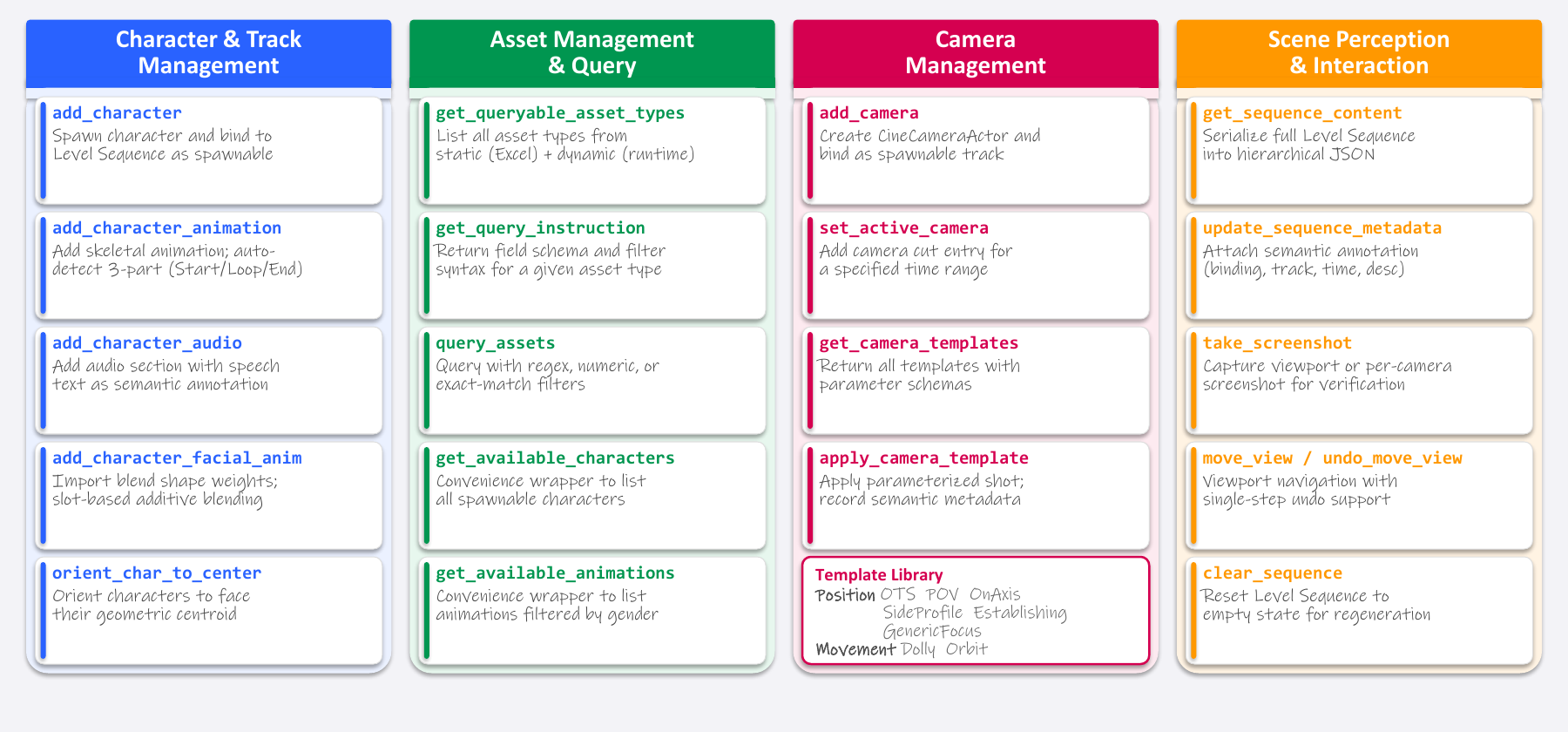}
    \caption{Overview of the Cutscene Toolkit's four functional modules and their MCP tool interfaces.}
    \label{fig:toolkit_overview}
\end{figure}

\subsubsection{Character \& Track Management}
\label{sec:toolkit_character}

This module manages the complete lifecycle of characters within a Level Sequence. At the \textit{binding} level, characters are instantiated from project asset data and registered as spawnable tracks in the sequencer. At the \textit{section} level, the module provides tools to populate each character's tracks with body animations, audio clips, and facial expressions. A central design principle is the loader/receiver abstraction: the toolkit invokes project-registered loader functions (decorated with \code{@register_loader}) for character instantiation, allowing different projects to customize how character blueprints are spawned and bound without modifying core toolkit code.

The exposed tools are:

\begin{itemize}[nosep]
    \item \code{add_character()} --- Spawns a character actor via the registered loader and binds it to the Level Sequence as a spawnable track.
    \item \code{orient_character_to_center()} --- Computes the geometric centroid of the specified characters and orients each to face it, for natural multi-character dialogue staging.
    \item \code{add_character_animation()} --- Adds skeletal animation sections to a character's animation track.
    \item \code{add_character_audio()} --- Adds an audio section to a character's audio track, with optional speech text stored as a semantic annotation for downstream reference.
    \item \code{add_character_facial_animation()} --- Adds skeletal animation sections to a character's dedicated facial animation track.
\end{itemize}

\subsubsection{Asset Management \& Query}
\label{sec:toolkit_asset}

Cutscene production requires access to diverse assets---character blueprints, skeletal animations, audio clips, facial expression data---each with rich metadata that agents must query to make informed decisions. The asset management module provides a dual-layer architecture that unifies two fundamentally different asset lifecycles under a single query interface.

\paragraph{Static Assets.} Pre-authored assets (characters, animations, etc.) are stored in structured Excel spreadsheets with a four-row header schema: \textit{identifier} (unique key), \textit{loader} (which registered function handles this asset type), \textit{public data} (metadata visible to LLM agents, such as names, descriptions, and semantic tags), and \textit{private data} (engine-internal information like asset paths and blueprint references, invisible to agents). This public/private separation is a deliberate design choice: it prevents LLMs from hallucinating or misusing raw engine paths while providing them with semantically meaningful information for decision-making.

\paragraph{Dynamic Assets.} Assets generated at runtime---such as TTS-synthesized speech audio or audio-driven facial animation data---are ingested through the public \code{import_dynamic_asset} tool, which supports three data-source modes: \textit{base64} for encoded binary payloads (suited for cross-machine delivery), \textit{file\_path} for local files already on disk, and \textit{url} for network resources . Each import request specifies a \texttt{data\_type} (an \textit{importable type} such as \texttt{audio\_wav}) that selects the appropriate registered receiver (\code{@register_receiver}) for format-specific processing (e.g., WAV$\rightarrow$Sound Wave import, NPZ$\rightarrow$Animation Asset conversion). The toolkit assigns a normalized identifier following a \texttt{\{prefix\}\_\{hint\}\_\{hex\_suffix\}} scheme that guarantees global uniqueness and human readability. Once processed, the record is persisted in a JSON registry and becomes queryable through the same interface as static assets.

The module exposes two categories of tools.

\paragraph{Query Tools.} These tools follow a progressive-disclosure pattern for asset retrieval:

\begin{itemize}[nosep]
    \item \code{get_queryable_asset_types()} --- Returns all available asset types from both static (Excel sheets) and dynamic (runtime) sources.
    \item \code{get_query_instruction()} --- Returns the field schema and filter syntax guide for a specific asset type, enabling agents to construct well-formed queries.
    \item \code{query_assets()} --- Queries assets with optional filtering via a lightweight DSL supporting regex (\texttt{/pattern/}), numeric comparisons (\texttt{>=N}, \texttt{<N}), and exact match.
    \item \code{get_available_characters()} / \code{get_available_animations()} --- Convenience wrappers for the most frequently queried asset types.
\end{itemize}

\paragraph{Import Tools.} These tools provide a progressive-discovery pattern for dynamic asset ingestion:

\begin{itemize}[nosep]
    \item \code{get_importable_asset_types()} --- Returns all registered importable types with descriptions and supported file extensions.
    \item \code{get_import_guide()} --- Returns recommended metadata fields and import behavior for a specific type.
    \item \code{import_dynamic_asset()} --- Imports a dynamic asset from any supported data source and returns the assigned identifier.
\end{itemize}

\subsubsection{Camera Management}
\label{sec:toolkit_camera}

The camera module follows a layered design. The low-level API handles camera actor creation and cut-track management, while the high-level template system parameterizes cinematic shot language---agents specify intent (e.g., ``over-the-shoulder from Alice to Bob'') rather than raw transforms. Template positions are computed automatically from character skeletal bone positions (head, spine, shoulder), making shots adapt naturally to characters of different scales and proportions. Crucially, template-generated positions serve as \textit{informed initial placements} rather than final results: when finer compositional control is desired, a vision-capable sub-agent can enter a visual refinement loop---iteratively invoking viewport navigation (\code{move_view}) and screenshot capture tools (Section~\ref{sec:toolkit_perception}) to evaluate framing and adjust the camera pose with closed-loop visual feedback. This two-stage workflow---semantic template placement followed by perception-driven refinement---combines the efficiency of parameterized templates with the precision of visual grounding. Correspondingly, the module exposes a compact set of tools that cover camera creation, shot activation, template discovery, and template application.

\begin{itemize}[nosep]
    \item \code{add_camera()} --- Creates a \texttt{Cine\-Camera\-Actor} and binds it to the Level Sequence as a spawnable.
    \item \code{set_active_camera()} --- Adds a camera cut entry to the sequence's Camera Cut track for the specified time range.
    \item \code{get_available_camera_templates()} --- Returns structured JSON descriptions of all available templates with their parameter schemas.
    \item \code{apply_camera_template()} --- Applies a parameterized shot template and records its semantic description as sequence metadata.
\end{itemize}

\noindent The template library comprises six \textit{position templates} and two \textit{movement templates}:

\begin{itemize}[nosep]
    \item \textbf{Position:} OTS (Over-the-Shoulder, with near/mid/high variants), POV (Point-of-View), OnAxis (midpoint between two actors), SideProfile (left/right at bone level), Establishing (wide two-actor shot), and GenericFocus (flexible single-target with configurable distance, pitch, yaw, and bone attachment).
    \item \textbf{Movement:} Dolly (approach/retreat by ratio) and Orbit (arc around look-at target by angle).
\end{itemize}

\begin{figure}[htbp]
    \centering
    \includegraphics[width=\textwidth]{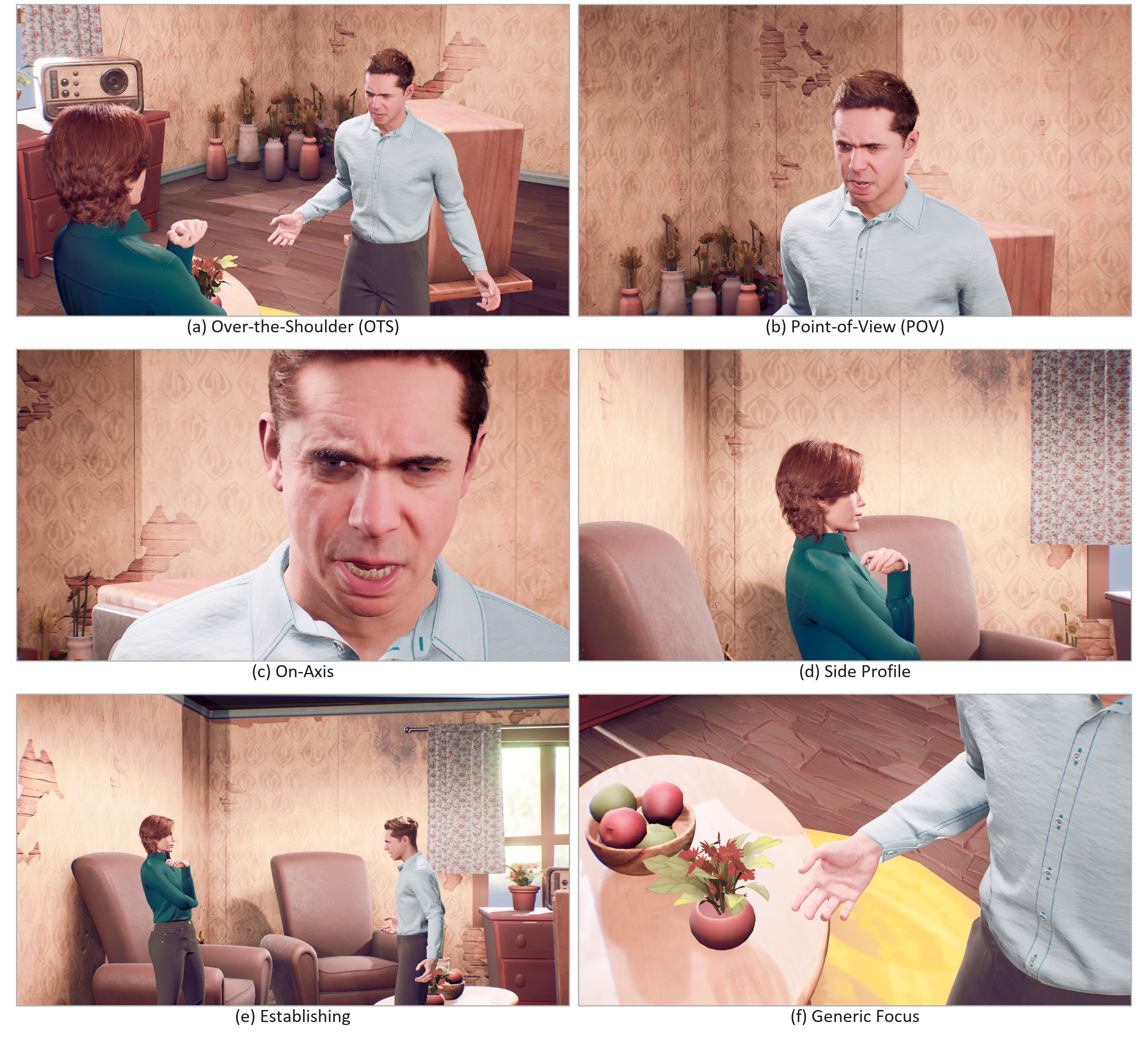}
    \caption{Camera template library provided by the Cutscene Toolkit. See Appendix~\ref{app:camera_specs} for more details.}
    \label{fig:camera_templates}
\end{figure}

\subsubsection{Scene Perception \& Interaction}
\label{sec:toolkit_perception}

Effective agent decision-making requires awareness of the current sequence state. This module provides three categories of functionality: state serialization for reading the sequence content as structured data, semantic metadata management for preserving authorial intent across tool calls, and editor interaction for visual verification through viewport control and screenshot capture.

\begin{itemize}[nosep]
    \item \code{get_sequence_content()} --- Serializes the entire Level Sequence into a hierarchical JSON structure, including all character bindings, their animation/audio/facial tracks with per-section timing and asset references, camera cut entries, and embedded semantic metadata. This is the primary mechanism for \textit{automatic state synchronization} (Section~\ref{sec:multiagent}).
    \item \code{update_sequence_metadata()} --- Attaches a semantic annotation block (binding name, track, time range, description) to the sequence, enabling agents to record intent (e.g., ``OTS shot from Alice to Bob'') that persists through subsequent queries.
    \item \code{take_editor_screenshot()} / \code{take_camera_screenshot()} --- Captures viewport or camera-specific screenshots for visual verification of the current scene state.
    \item \code{move_view()} / \code{undo_move_view()} --- Provides character-style viewport navigation (forward/horizontal constrained to XY plane, vertical along global Z) with single-step undo, enabling agents to inspect the scene from different angles.
\end{itemize}

\subsubsection{Implementation Details}
\label{sec:toolkit_impl}

The toolkit operates as a three-layer architecture: an MCP server exposes tool definitions over SSE transport; a Task Scheduler mediates execution on Unreal Engine's game thread; and the UE Python API layer performs the actual engine operations.

\noindent\textbf{Main Thread Scheduling.} Since the MCP server's blocking event loop must run on a dedicated sub-thread, a \code{@schedule_execute_in_main_thread} decorator routes all UE API side-effects back to the game thread via Slate post-tick callbacks, ensuring thread-safe access to engine state. Because each tool invocation is itself blocking, tools that depend on rendered output (e.g., screenshot capture) require intermediate frames to be flushed; a generator-based \texttt{yield} mechanism lets such tools suspend execution for a specified number of ticks, guaranteeing the render pipeline completes before the tool reads its results.

\noindent\textbf{Modular Tool Registration.} \texttt{MCP\-Tool\-Registry} enables tools to be defined across multiple Python modules with \texttt{@tool()} decorators. All registrations are deferred until the FastMCP application is instantiated, at which point \code{apply_to_app()} registers them in a single pass. This decouples tool definition from server lifecycle.

\noindent\textbf{Extensible Asset Loaders.} The \code{@register_loader} and \code{@register_receiver} decorator system allows project-specific customization layers (e.g., \code{cutscene_provider_custom/}) to override default asset handling---different character blueprint pipelines, animation import formats, or facial data processing---without modifying the core toolkit. The asset system explicitly decouples three type concepts: \textit{importable type} (governing import validation and receiver dispatch, e.g., \texttt{audio\_wav}), \textit{loader type} (governing runtime loading into the Level Sequence, e.g., \texttt{audio}), and \textit{asset kind} (the semantic category visible to agents and the query interface, e.g., \texttt{Audio}). A unified \texttt{AssetRecord} Pydantic model provides a consistent runtime representation for both static (Excel-sourced) and dynamic (runtime-imported) assets, while \texttt{LoaderCallable} and \texttt{ReceiverCallable} Protocol types enforce strict function signatures across all registered handlers.

\noindent\textbf{Parameter Validation.} Pydantic models define structured schemas for complex inputs (e.g., camera template arguments). The JSON schemas are automatically exposed to agents through template query tools, providing self-documenting parameter specifications.

\noindent\textbf{Unified Logging.} A centralized logging system (\code{get_logger}) routes all Python-side log output through UE's native log framework via a C++ bridge, integrating with UE's built-in log filtering and output facilities.


\subsection{Agent System}
\label{sec:multiagent}

Built upon the Cutscene Toolkit, we implement an \textbf{Agent System} that drives automated cutscene generation. Two core challenges motivate its design: (1)~managing complex prompts within limited context windows so that the agent simultaneously understands domain-specific rules \emph{and} perceives real-time engine state; and (2)~orchestrating flexible workflows through on-demand subagent delegation and visual reasoning feedback to achieve high-quality outputs. The complete architecture diagram is provided in Figure~\ref{fig:architecture} (Appendix~\ref{app:agent_specs}).

\subsubsection{Prompt and Context Management}
\label{sec:prompt_management}

The quality of agent outputs depends critically on prompt engineering. We design a composable, priority-driven, token-budget-aware prompt management system that assembles system prompts dynamically and maintains coherent context across long generation sessions.

\paragraph{Composable Prompt Architecture.}
System prompts are built from a hierarchy of \texttt{Prompt\-Element} objects, each carrying a numeric \textit{priority} and a \code{get_token_count()} method backed by \texttt{tiktoken}. Concrete element types include \texttt{System\-Instruction} (core role and safety rules; highest priority), \texttt{Context\-Block} (XML-tagged knowledge blocks), and \texttt{Text\-Element} (free-form guidance).

The \texttt{Prompt\-Manager} assembles the final prompt by: (i)~collecting all registered elements, (ii)~sorting by descending priority, (iii)~greedily selecting elements until the token budget is exhausted, and (iv)~rendering selected elements in category order (instructions first, then context blocks). This mechanism enables graceful degradation under context-window pressure: low-priority auxiliary information is automatically pruned while high-priority core rules are always retained.

Furthermore, the toolkit backend can define project-specific prompt fragments---such as character naming conventions, scene constraints, or art style guidelines---which are automatically pulled via MCP at agent initialization and injected into the system prompt. This allows the same Agent System to adapt to different project contexts without modifying agent-side code.

\paragraph{Automatic State Injection.}
Before each LLM reasoning step, the ~\code{get_sequence_content} tool is automatically invoked to retrieve the complete Level Sequence state as hierarchical JSON---including all character bindings, animation/audio/facial tracks with per-section timing, camera cut entries, and semantic metadata---and injects it into the conversation context as an XML-wrapped block:

\begin{verbatim}
<current_cutscene_content>
  Contents in current cutscene:
  { "bindings": [...], "toplevel_tracks": [...], ... }
</current_cutscene_content>
\end{verbatim}

This ensures the agent perceives the full cutscene state at every decision point, achieving state consistency without explicit inter-process messaging. Unlike traditional unidirectional function calling---where the agent only receives individual tool return values and has no awareness of cumulative state changes---our bidirectional approach automatically synchronizes global state after every interaction.

\paragraph{Tool Call and History Compression.}
In extended generation sessions, tool call history grows rapidly and can dominate the context window. We employ a category-aware compression strategy that differentiates tools by their information-retention characteristics:

\begin{itemize}
    \item \textbf{Mutation tools} (e.g., \code{add_character}, \code{add_character_animation}) modify the Level Sequence directly. Since their cumulative effects are fully captured by the automatically injected cutscene state JSON, these calls are \textit{compressed first}---replaced by a concise XML summary listing only tool names---with minimal information loss.
    
    \item \textbf{Query tools} (e.g., \code{query_assets}, \code{get_available_animations}) retrieve asset catalogs that the agent references throughout the session. The system retains the \textit{most recent invocation} of each query tool, ensuring the agent always has access to up-to-date asset information without redundant re-queries.
\end{itemize}

Only the \(N\) most recent tool calls are kept in full detail (arguments and return values). Compared to uniform threshold-based compression, this category-aware policy preserves critical query context while more aggressively reclaiming token budget from mutation calls whose effects are already observable in the injected state.

\subsubsection{Workflow and Subagent Delegation}
\label{sec:subagent}

The main agent's execution loop is built on the OpenAI Agents SDK with customized substep iteration: inject cutscene state $\rightarrow$ LLM reasoning $\rightarrow$ tool execution $\rightarrow$ history compression $\rightarrow$ completion check, with streaming event output for real-time interaction via CLI or web UI.

\paragraph{Subagent Delegation.}
We design a hierarchical subagent framework that enables the main agent to autonomously create, dispatch, and manage specialized subagents during generation. The main agent invokes subagents through the \code{run_subagent} tool, which supports two complementary modes: \textit{preset templates} for recurring domain tasks and \textit{custom mode} for ad-hoc delegation.

\begin{itemize}
    \item \textbf{Preset specialist subagents.}\quad
Drawing inspiration from the division of labor in professional film production, we position the main agent as a \textit{director} who orchestrates a crew of domain specialists. Each preset template defines a self-contained subagent profile comprising three components: (1)~a tailored system prompt encoding domain expertise and behavioral guidelines; (2)~a tool whitelist restricting the subagent to only the MCP tools relevant to its specialty; and (3)~a maximum turn budget bounding execution length. A complete listing of all preset subagents and their configurations is provided in Appendix~\ref{app:agent_specs}. A few representative subagents include:

\begin{itemize}
    \item \textit{Animation Specialist} --- responsible for selecting and timing character performance animations based on the script's emotional beats and stage directions; restricted to animation query and assignment tools.
    \item \textit{Cinematographer} --- controls camera placement, shot composition, and lens parameters following established cinematographic conventions; restricted to camera creation and template application tools.
    \item \textit{Sound Designer} --- handles text-to-speech generation, audio track attachment, and audio-to-facial-animation synchronization for each character's dialogue lines, ensuring precise lip-sync alignment; restricted to TTS, audio, and facial animation tools.
\end{itemize}

Each subagent operates with a fully independent context window: it receives the current Level Sequence state upon creation but maintains its own conversation history, preventing cross-contamination of reasoning across different creative domains. 

\item \textbf{Dynamic custom subagents.}\quad
Beyond preset templates, the main agent possesses the ability to dynamically construct subagents at runtime. When encountering a subtask not covered by existing presets, the main agent specifies a custom subagent by providing: \code{custom_instructions} (free-form task description and behavioral constraints), \code{custom_tool_scope} (an ad-hoc tool whitelist), and an execution turn limit. The custom subagent's full lifecycle---creation, context initialization, execution, and result collection---is managed entirely by the main agent, requiring no human intervention. This mechanism effectively allows the agent system to \textit{extend its own capability surface} on the fly.

\item \textbf{Isolation and coordination.}\quad
All subagents---whether preset or custom---are sandboxed via MCP tool filtering that exposes only whitelisted tools, preventing unintended side effects on the Level Sequence. Upon completion, each subagent returns a structured JSON summary (status, tool call count, result description) to the main agent, which uses this feedback to decide subsequent actions. This design yields several advantages: it \textit{decomposes} a monolithic long-horizon task into manageable subtasks with bounded context, \textit{reduces} the risk of error propagation by isolating creative decisions, and \textit{improves} generation quality by allowing each specialist to reason within a focused, domain-specific context.

\end{itemize}

\subsubsection{Visual Reasoning Feedback Loop}
\label{sec:visual_feedback}

A fundamental limitation of existing LLM-based content generation systems is that the agent operates \textit{blind}: it emits tool calls to construct scenes but has no capacity to observe the visual outcome of its actions. This is analogous to a film director giving instructions with their eyes closed---the agent may produce structurally valid sequences that are nonetheless aesthetically poor due to undetectable issues such as awkward character framing, occluded compositions, or misaligned spatial layouts. Prior works~\cite{xu2025filmagent} partially address this through scripted evaluation heuristics, but lack the ability to \textit{perceive and react to} the actual rendered result in real time.

We address this gap by introducing a closed-loop visual reasoning mechanism that grants the agent the ability to \textit{see} the current rendering result of the cutscene within the engine and autonomously refine its creative decisions based on visual feedback. This constitutes, to our knowledge, the first integration of visual perception into an agent-driven game engine content pipeline, and is the core enabler of truly bidirectional agent--engine interaction.

\paragraph{Perceive--Reason--Act Loop.}
The feedback mechanism is structured as an iterative \textit{Perceive $\rightarrow$ Reason $\rightarrow$ Act} cycle, illustrated as follows:

\noindent\textbf{Perceive.}\quad The agent invokes \code{take_editor_screenshot} via MCP to capture a screenshot of the current Unreal Engine editor viewport. The resulting image is encoded and injected into the multimodal LLM's context as a visual input alongside the textual cutscene state. This gives the agent a pixel-level observation of exactly what the cutscene looks like at the current point in time---including character spatial relationships, camera framing, lighting conditions, and overall composition.

\noindent\textbf{Reason.}\quad The vision-language model analyzes the captured image against the intended creative goals (derived from the input script, storyboard, or the agent's internal plan). The model evaluates multiple quality dimensions---shot composition, rule-of-thirds adherence, character visibility, depth-of-field appropriateness, and narrative expressiveness---and produces a structured diagnostic identifying specific deficiencies (e.g., ``character is partially occluded by foreground geometry'' or ``camera angle does not convey the intended emotional intensity'').

\noindent\textbf{Act.}\quad Based on the diagnostic, the agent issues corrective tool calls: \code{move_view} to adjust camera position and rotation, \code{apply_camera_template} to switch to a more suitable shot type, or character management tools to reposition actors. The cycle then repeats---the agent captures a new screenshot, re-evaluates, and continues refining until the composition satisfies quality criteria or a maximum iteration budget is reached.

This loop transforms the agent from a one-shot generator into an \textit{iterative refinement engine} that converges toward higher-quality outputs through self-correction---mirroring the workflow of a human cinematographer who adjusts the camera while monitoring a viewfinder.

\paragraph{Generalization to Broader Perception Tasks.}
The visual reasoning mechanism is not limited to camera refinement. Through the custom subagent mode (Section~\ref{sec:subagent}), any vision-language model can be deployed as a perception agent for diverse review tasks: scene layout validation (verifying character placement matches stage directions), animation quality assessment (checking whether selected animations convey the intended emotions), and continuity checking (detecting visual inconsistencies across shots). This modularity compensates for the inherent visual blindness of text-only LLMs by introducing dedicated perception stages at critical points in the generation pipeline.

Additionally, the system provides a \code{video_understanding} tool that applies a visual model to analyze reference videos supplied as input. The model extracts cinematographic patterns---shot types, transition rhythms, camera movement dynamics, and pacing structures---and encodes them as structured textual guidance for the generation agent, enabling \textit{reference-driven} cutscene creation.

\subsection{Key Technical Challenges}
\label{sec:challenges}

Developing an automated cutscene generation system that operates within a game engine involves addressing several non-trivial technical challenges that are often underestimated in prior work. We discuss key difficulties encountered and our approaches to resolving them.

\subsubsection{Spatial Reasoning and Character Positioning}

Current LLMs exhibit weaknesses in numerical spatial reasoning. When asked to specify 3D coordinates directly (e.g., \texttt{location: [100.0, -50.0, 0.0], rotation: [0.0, 0.0, 90.0]}), models frequently produce degenerate configurations---characters placed hundreds of units apart for an intimate dialogue, actors spawned outside scene boundaries, or facing away from conversation partners. Prior approaches mitigate this by confining characters to predefined stage marks~\cite{xu2025filmagent}, but such constraints sacrifice creative flexibility and produce formulaic blocking. We address this through two complementary strategies:

\noindent\textbf{Semantic Spatial Abstraction.} Rather than exposing raw coordinates, we design spatial parameters as high-level semantic interfaces that the engine resolves into precise values. For orientation, tools accept descriptors such as \code{"face_north"}, \code{"face_character:Alice"}, or \code{"turn_left_45"} instead of Euler angles, which the MCP server translates into exact rotations on the engine side. For positioning, characters can be placed relative to scene anchors (e.g., ``2 meters in front of Alice'') rather than in absolute world coordinates. The API further enforces project-defined bounding volumes and default height offsets, preventing out-of-bounds placement and ground-plane violations. Together, these abstractions reduce spatial specification to directional language, relative distances, and named references---a reasoning level LLMs handle reliably---while the engine backend performs the numerical translation.

\noindent\textbf{Visual Perception--Correction Loop.} Semantic abstraction eliminates most positioning failures but cannot guarantee aesthetically satisfactory staging in all cases---subtle issues like character occlusion or awkward spacing require visual judgment. We therefore leverage the visual reasoning feedback loop (Section~\ref{sec:visual_feedback}) as a second line of defense: after initial placement, a vision-capable subagent captures a viewport screenshot, evaluates the arrangement, and issues corrective repositioning commands. This perceive--reason--act cycle enables the system to converge on visually coherent staging even from imperfect initial conditions.

\subsubsection{Environment Awareness and Collision Avoidance}

The agent system lacks inherent understanding of 3D environment geometry. Without explicit perception of scene obstacles, the system cannot prevent characters from being placed inside walls, furniture, or terrain features, nor avoid character-to-character interpenetration during animations. Camera trajectories may likewise intersect scene geometry. These visual artifacts would be immediately flagged by human artists but remain invisible to text-only agents operating without geometric feedback.

We address this with a generate-then-validate strategy. A suite of collision-detection functions is exposed through the MCP server, querying the Unreal Engine physics subsystem to perform overlap tests between character capsules, skeletal mesh bounds, and static-environment geometry and returning structured collision reports. After the primary generation agents complete a cutscene, a dedicated \emph{validation subagent} scrubs through the entire performance timeline: at each sampled keyframe it invokes the collision-detection toolkit, identifies interpenetration events, and issues corrective repositioning or animation adjustments to resolve them while preserving the original directorial intent.

\subsubsection{Multi-Track Timing Synchronization}

Cutscenes require precise coordination across multiple temporal tracks---body animation, facial expression, dialogue audio, and camera cuts. Errors accumulate across sequential tool calls, easily leading to audio-visual desynchronization, animations extending beyond shot boundaries, or gaps and overlaps in continuous action sequences.

We mitigate this at multiple levels. Every track-manipulation tool is designed to return precise timestamps (e.g., the exact time interval of an inserted animation or audio clip), so that subsequent calls can chain from authoritative values rather than LLM estimates. As the cutscene is progressively assembled, a structured content record maintained by the \code{get_sequence_content} tool tracks the current state of all sections and their timing across every track. Once the primary generation is complete, dedicated \emph{validation subagents} scan this final content record for synchronization anomalies---such as misaligned audio-lip tracks, missing facial-expression segments, or animation gaps---and dispatch the corresponding specialized subagents to apply targeted corrections. This pipeline effectively reduces track desynchronization and coverage omissions without burdening the generation agents with exhaustive cross-track bookkeeping.

\section{Evaluation Framework}
\label{sec:evaluation}

Evaluating automatically generated cutscenes presents unique challenges that are not fully addressed by existing benchmarks in either the tool-use or the video generation literature. In this section, we introduce CutsceneBench, a hierarchical evaluation framework specifically designed for agent-driven cutscene generation. We describe the design rationale behind each evaluation layer, detail the benchmark construction process, and report evaluation results across a range of LLMs.

\subsection{Evaluation Challenges}
\label{sec:eval_challenges}

Agent-driven cutscene generation occupies an unusual position at the intersection of tool-use evaluation and creative content assessment. Several properties of this task make direct application of existing benchmarks insufficient:

\begin{enumerate}
    \item \textbf{Absence of Canonical Ground Truth.} Unlike function-calling benchmarks such as BFCL~\cite{patil2025bfcl} where expected outputs can be precisely specified, cutscene generation admits multiple valid solutions. The same script can be realized through different camera placements, animation selections, and timing choices---all of which may be equally acceptable.
    
    \item \textbf{Long-Horizon Multi-Step Dependencies.} A typical cutscene requires 10--60 sequential tool calls with strict inter-call dependencies (e.g., a character must be spawned before animations can be assigned). This is substantially more complex than the single-step or short-horizon tasks evaluated by most tool-calling benchmarks~\cite{xu2023toolllm, lu2024toolsandbox}.
    
    \item \textbf{Stateful Side Effects.} Each tool call modifies the persistent state of the Unreal Engine Level Sequence. Evaluation must account for the cumulative effect of the entire call trajectory on the final artifact, not just the correctness of individual calls in isolation.
    
    \item \textbf{Multi-Dimensional Quality.} Cutscene quality spans technical correctness (valid tool calls, proper dependencies), structural integrity (well-formed sequences), and subjective creative quality (cinematography, narrative coherence)---dimensions that require fundamentally different evaluation methodologies.
\end{enumerate}

These challenges motivate a layered evaluation design that addresses each dimension with appropriate methods, progressing from fully automated low-level verification to higher-level quality assessment.

\subsection{CutsceneBench: Hierarchical Evaluation Architecture}
\label{sec:cutscenebench}

\begin{figure}[htbp]
    \centering
    \includegraphics[width=\textwidth]{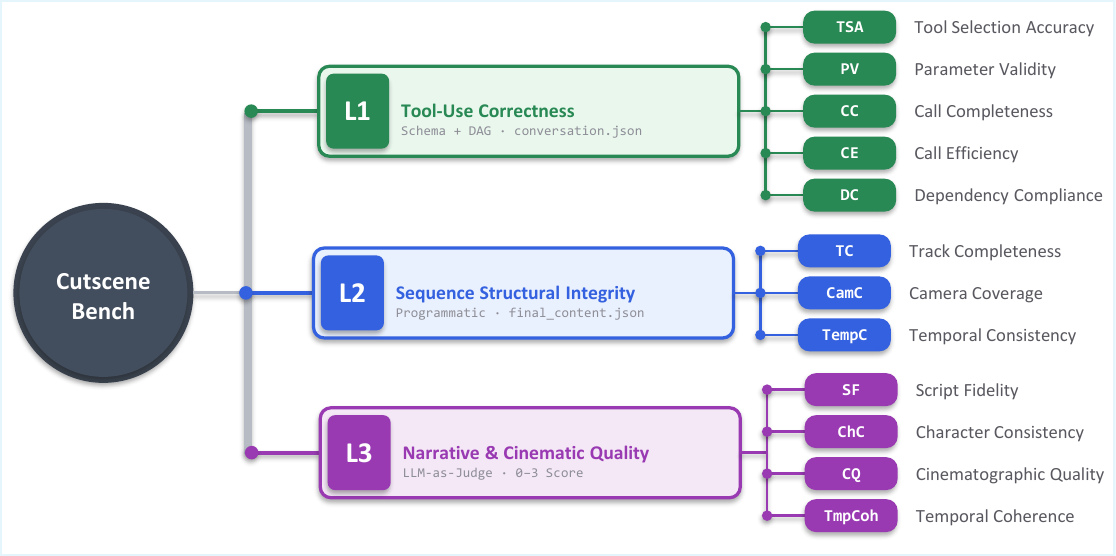}
    \caption{CutsceneBench hierarchical evaluation architecture. Three layers progressively assess generation quality: L1 validates tool-call correctness and inter-tool dependency ordering, L2 programmatically inspects the final Level Sequence structure, and L3 evaluates narrative and cinematic quality via LLM-as-Judge.}
    \label{fig:eval_framework}
\end{figure}

We propose a three-layer hierarchical evaluation framework, illustrated in Figure~\ref{fig:eval_framework}. The three layers form a progressive evaluation pipeline that mirrors the hierarchy of concerns in cutscene production. Layer~1 (Tool-Use Correctness) operates at the \emph{atomic level}, evaluating whether each individual tool call is well-formed: correct tool selection, valid parameters, and proper inter-call dependency ordering. High L1 scores are a necessary precondition for any model to be considered capable of cutscene generation---a model that cannot reliably produce correct function calls cannot produce a viable cutscene. Layer~2 (Structural Integrity) shifts focus from individual calls to the \emph{holistic artifact}, assessing whether the generated Level Sequence is structurally complete and temporally coherent as a whole---i.e., whether the cumulative effect of the tool-call trajectory yields a well-formed production asset. Layer~3 (Narrative and Cinematic Quality) addresses the highest-level concern: \emph{aesthetic quality}. Given a structurally sound sequence, does it realize the input script with effective cinematography, coherent pacing, and professional visual storytelling? This progression---from atomic correctness to structural completeness to aesthetic merit---enables both fine-grained failure diagnosis and aggregate quality scoring, with each layer independently applicable.

\subsubsection{Layer 1: Tool-Use Correctness}

The foundation of our evaluation framework assesses whether the agent's tool-call trajectory is well-formed: individual calls must be semantically appropriate with valid parameters, essential operations must be covered without redundancy, and strict inter-tool ordering constraints must be respected. This layer operates entirely on the recorded tool-call trajectory and requires no engine execution. Inspired by the AST-based evaluation of BFCL~\cite{patil2025bfcl} and the Tool Graph formulation of TaskBench~\cite{shen2024taskbench}, we define five metrics:

\begin{itemize}
    \item \textbf{Tool Selection Accuracy (TSA)}: The proportion of tool calls that invoke the correct tool for the intended operation.
    \[
    \text{TSA} = \frac{|\{\text{calls with correct tool selection}\}|}{|\{\text{total tool calls}\}|}
    \]
    
    \item \textbf{Parameter Validity (PV)}: The proportion of tool calls with parameters satisfying type constraints, value ranges, and referential integrity (e.g., character names must reference previously spawned characters). Checked via schema validation against the Pydantic models defined in our MCP toolkit.
    \[
    \text{PV} = \frac{|\{\text{calls with valid parameters}\}|}{|\{\text{total tool calls}\}|}
    \]
    
    \item \textbf{Call Completeness (CC)}: The recall of essential operations. Given a script, certain tool calls are necessary---e.g., each dialogue line requires TTS generation, audio attachment, and facial animation. We measure the fraction of such required operations that are actually performed.
    \[
    \text{CC} = \frac{|\{\text{essential operations performed}\}|}{|\{\text{essential operations required}\}|}
    \]
    
    \item \textbf{Call Efficiency (CE)}: The absence of redundant or unnecessary tool calls. Redundant calls waste tokens and may introduce unintended side effects.
    \[
    \text{CE} = 1 - \frac{|\{\text{redundant calls}\}|}{|\{\text{total tool calls}\}|}
    \]

    \item \textbf{Dependency Compliance (DC)}: Cutscene generation imposes strict ordering constraints---e.g., \code{add_character_animation} requires a prior \code{add_character} for the same entity; \code{set_active_camera} requires a prior \code{add_camera}. We model these inter-tool dependencies as a directed acyclic graph (DAG) $G = (V, E)$ and verify that every dependency edge is respected in the agent's call trajectory $\tau$:
    \[
    \text{DC} = 1 - \frac{|\{\text{dependency violations in } \tau\}|}{|\{\text{dependency edges applicable to } \tau\}|}
    \]
\end{itemize}

\begin{figure}[htbp]
    \centering
    \includegraphics[width=\textwidth]{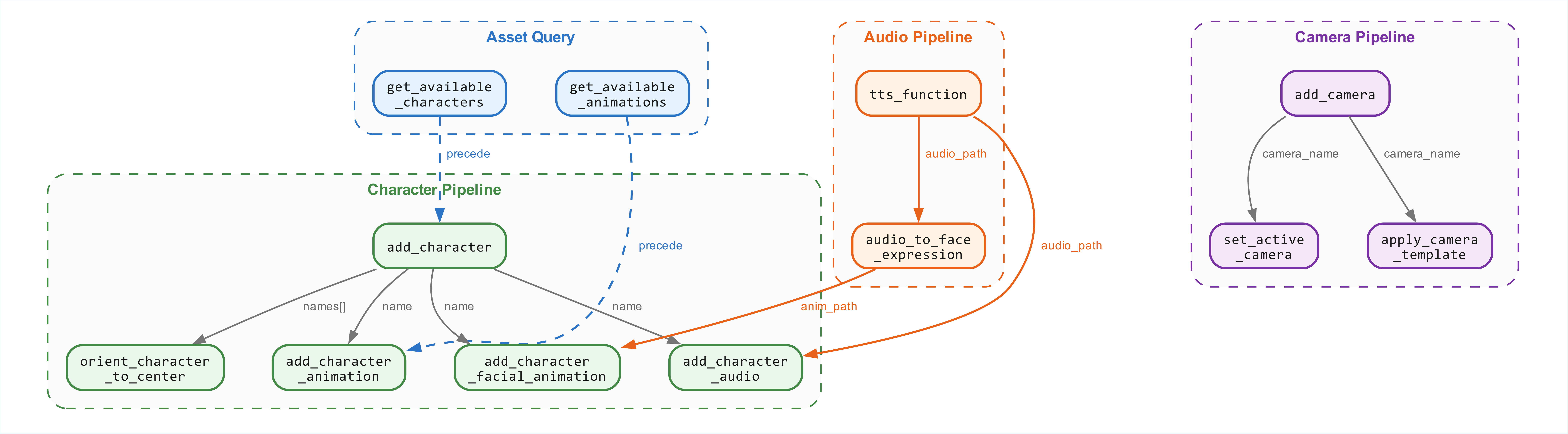}
    \caption{Inter-tool dependency DAG for the Cutscene Toolkit. Nodes are grouped by functional pipeline: \textcolor[HTML]{1565C0}{Asset Query} (blue), \textcolor[HTML]{2E7D32}{Character Pipeline} (green), \textcolor[HTML]{E65100}{Audio Pipeline} (orange), and \textcolor[HTML]{6A1B9A}{Camera Pipeline} (purple). Dashed edges denote precedence constraints (query before mutation); solid edges denote instance-level parameter dependencies. The DC metric verifies that the agent's trajectory respects this ordering.}
    \label{fig:dependency_dag}
\end{figure}

The dependency graph for our toolkit is illustrated in Figure~\ref{fig:dependency_dag}.

A key advantage of the DC metric is that it requires no ground truth---the dependency graph is derived from the toolkit's API contracts and can be validated purely through topological analysis of the call trajectory.

\subsubsection{Layer 2: Sequence Structural Integrity}

Beyond the tool-call trajectory, the generated Level Sequence must be structurally well-formed as a whole. This layer performs programmatic validation on the final sequence state, analogous to the state-based evaluation approach of $\tau$-bench~\cite{sierra2024tau}---we evaluate the end state rather than the trajectory that produced it.

We define three sub-metrics, each producing a continuous score in $[0,1]$:

\begin{itemize}
    \item \textbf{Track Completeness (TC)}: Every character referenced in the script must have a corresponding binding in the Level Sequence with the requisite sub-tracks (skeletal animation, audio, facial animation). For each character~$c$ in the expected set~$\mathcal{C}$, we verify the existence of each required track type~$t$ and check that it contains at least one non-empty section. When a ground-truth sequence is available, only track types present in the GT are required (e.g., non-speaking characters are not penalized for missing audio/facial tracks):
    \[
    \text{TC} = \frac{\sum_{c \in \mathcal{C}} \sum_{t \in \mathcal{T}(c)} \mathbb{1}[\text{track } t \text{ exists and is non-empty}]}{|\{(c,t) \mid c \in \mathcal{C},\; t \in \mathcal{T}(c)\}|}
    \]
    where $\mathcal{T}(c)$ denotes the set of required track types for character~$c$.

    \item \textbf{Camera Coverage (CamC)}: The Camera Cut track must span the entire effective sequence duration~$D$ without gaps. We merge all camera cut intervals and compute the fraction of $D$ that is covered:
    \[
    \text{CamC} = \min\!\left(\frac{\sum_{i} |I_i \cap [0, D]|}{D},\; 1\right)
    \]
    where $\{I_i\}$ are the merged camera cut intervals clipped to $[0, D]$. Any uncovered interval constitutes a gap during which no active camera is assigned.

    \item \textbf{Temporal Consistency (TempC)}: Within each character's tracks, animation and audio sections must not overlap beyond a tolerance threshold $\epsilon$. Additionally, each audio section must align temporally with a corresponding facial animation section (start and end times within tolerance $\delta$). We count the total number of pairwise checks and the number of violations:
    \[
    \text{TempC} = 1 - \frac{|\{\text{temporal violations}\}|}{|\{\text{temporal checks}\}|}
    \]
    where violations include intra-track overlaps (consecutive sections $s_i, s_{i+1}$ with $\text{end}(s_i) > \text{start}(s_{i+1})\allowbreak + \epsilon$) and audio--facial misalignments exceeding~$\delta$.
\end{itemize}

\subsubsection{Layer 3: Narrative and Cinematic Quality}

The highest evaluation layer assesses subjective creative quality: does the generated cutscene faithfully realize the input script with professional-grade cinematography? This layer cannot be fully automated and requires either human evaluation or LLM-based assessment.

We adopt an LLM-as-Judge approach~\cite{zheng2023judging} with structured evaluation prompts, following best practices from G-Eval~\cite{liu2023geval}. A multimodal judge model independently evaluates each quality dimension via chain-of-thought reasoning, receiving the \emph{rendered cutscene video} alongside the original script as input. Each dimension is scored on a 0--25 integer scale, yielding an aggregate score of 0--100.

We evaluate four dimensions:

\begin{itemize}
    \item \textbf{Script Fidelity (SF)}: Whether all script elements---dialogue, character actions, emotional beats---are accurately represented in the rendered cutscene. Scoring: 0--25.
    
    \item \textbf{Character Consistency (ChC)}: Whether character identities, positions, and behaviors remain coherent throughout the cutscene. Scoring: 0--25.
    
    \item \textbf{Cinematographic Quality (CQ)}: Whether camera shot selection, composition, and cutting patterns follow established cinematographic conventions and serve the narrative effectively. Scoring: 0--25.
    
    \item \textbf{Temporal Coherence (TmpCoh)}: Whether the pacing, timing of cuts, and synchronization between audio, animation, and camera create a natural viewing experience. Scoring: 0--25.
\end{itemize}

The judge model produces a structured JSON evaluation containing both a discrete score and chain-of-thought reasoning for each dimension. We use a temperature of 0 to maximize reproducibility. The evaluation prompt explicitly instructs the judge to disregard environmental fidelity (e.g., test-map geometry versus described locations), focusing exclusively on agent-controlled elements: character performance, dialogue delivery, animation selection, and camera work. Due to the cost of multimodal evaluation, we sample 25 videos per model (5 per scenario tier S1--S5) for L3 assessment.

\subsection{Benchmark Construction}
\label{sec:benchmark_construction}

\subsubsection{Test Scenarios}

We construct a benchmark of test scenarios spanning varying levels of complexity, measured by character count, dialogue turns, and expected tool-call volume:

\begin{table}[h]
\centering
\begin{tabular}{@{}llcccc@{}}
\toprule
\textbf{ID} & \textbf{Description} & \textbf{Chars} & \textbf{Turns} & \textbf{Est. Duration} & \textbf{Complexity} \\
\midrule
S1 & Single character monologue & 1 & 1 & $\sim$20s & Simple \\
S2 & Two-person casual dialogue & 2 & 4 & $\sim$30s & Simple \\
S3 & Two-person emotional scene & 2 & 6 & $\sim$45s & Medium \\
S4 & Three-person conversation & 3 & 6 & $\sim$60s & Medium \\
S5 & Complex multi-turn scene & 3 & 10 & $\sim$90s & Complex \\
\bottomrule
\end{tabular}
\caption{CutsceneBench test scenarios. Complexity tiers are defined by the expected number of tool calls required for complete generation.}
\label{tab:test_scenarios}
\end{table}

Each scenario includes a script specifying characters, dialogue, stage directions, and intended emotional tone. Scripts are designed to exercise different aspects of the generation pipeline: S1 tests basic single-character workflow; S2--S3 test two-character dialogue with varying emotional complexity; S4 introduces multi-character spatial reasoning; and S5 stress-tests long-horizon planning and context management. For details of the scripts, please refer to Appendix~\ref{app:test_scenarios}.

\subsubsection{Ground Truth Strategy}

Constructing full ground-truth Level Sequences is prohibitively expensive, requiring professional technical artists to manually build reference sequences in Unreal Engine. We therefore adopt a tiered ground truth strategy that matches annotation cost to evaluation layer requirements:

\begin{itemize}
    \item \textbf{L1 (Low cost)}: For each scenario, we specify the expected set of essential tool calls and their dependency constraints. These annotations are text-based and can be constructed by developers familiar with the toolkit API, without requiring engine interaction.
    
    \item \textbf{L2 (Medium cost)}: Structural validation rules are defined once for the entire toolkit and applied universally. Per-scenario annotation is limited to specifying which characters and tracks should exist.
    
    \item \textbf{L3 (Variable cost)}: Narrative quality is assessed via multimodal LLM-as-Judge on rendered cutscene videos, requiring no per-scenario ground truth annotations. Cost scales with the number of sampled videos.
\end{itemize}

This tiered approach enables comprehensive evaluation while keeping annotation costs manageable.

\subsection{Evaluation Results}
\label{sec:eval_results}

We evaluate eight LLMs as the backbone for Cutscene Agent, including seven flagship-class models and one middle-size model. All models are evaluated under identical conditions: same system prompts, same MCP toolkit configuration, and same test scenarios across five complexity tiers (S1--S5). All evaluations were conducted in March 2026 using the model versions available at that time.

\subsubsection{Experimental Setup}

Models evaluated. We test the following eight LLMs: Claude Opus 4.6, Claude Sonnet 4.6, GPT-5.4, Qwen 3.5 Plus, Kimi K2.5, GLM-5, MiniMax M2.5, and Qwen 2.5-72B. The first seven are flagship-class models from their respective providers; Qwen 2.5-72B is included as a representative middle-size open-source model to probe the capability threshold for this task. All models are accessed via their respective API services.


\subsubsection{Layer 1: Tool-Use Correctness}

Table~\ref{tab:l1_results} reports tool-use correctness averaged across all 65 test scenarios per model.

\begin{table}[h]
\centering
\begin{tabular}{@{}lccccc@{}}
\toprule
\textbf{Model} & \textbf{TSA}$\uparrow$ & \textbf{PV}$\uparrow$ & \textbf{CC}$\uparrow$ & \textbf{CE}$\uparrow$ & \textbf{DC}$\uparrow$ \\
\midrule
Claude Opus 4.6      & 100.0 & 100.0 & 100.0 &  97.5 & 100.0 \\
Claude Sonnet 4.6    & 100.0 &  99.9 &  98.4 &  97.4 & 100.0 \\
GPT-5.4              & 100.0 &  96.6 &  95.7 &  97.4 &  98.5 \\
Qwen 3.5 Plus        &  99.9 &  97.1 &  94.5 &  99.7 &  99.5 \\
Kimi K2.5            &  99.3 &  97.4 &  91.8 &  98.6 &  98.7 \\
GLM-5                &  99.7 &  98.1 &  93.1 &  99.2 &  99.2 \\
MiniMax M2.5         &  99.5 &  91.6 &  90.9 &  98.7 &  99.2 \\
Qwen 2.5-72B         &  90.0 &  58.6 &  56.6 &  63.6 &  76.1 \\
\bottomrule
\end{tabular}
\caption{Layer 1 evaluation results averaged over all test scenarios. All values in \%.}
\label{tab:l1_results}
\end{table}

Among the seven flagship models, all achieve near-perfect Tool Selection Accuracy ($\geq$99.3\%) as expected. Dependency Compliance is also high overall ($\geq$98.5\%), though notably only Claude Opus 4.6 and Claude Sonnet 4.6 achieve a perfect 100.0\%---other models occasionally invoke mutation tools before querying available assets, violating the expected query-before-use ordering. The most discriminative metric is Call Completeness: Claude Opus 4.6 achieves perfect coverage, while other flagship models drop to 90--94\%, indicating more frequent omission of essential operations such as facial animation or audio attachment steps. Parameter Validity reveals meaningful differentiation: Claude Opus 4.6 achieves a perfect 100.0\%, while GPT-5.4 drops to 96.6\% and MiniMax M2.5 to 91.6\%---the latter two frequently fabricate character names or animation identifiers absent from the available asset library.

Middle-size model gap. Qwen 2.5-72B, the sole middle-size model in our evaluation, exhibits a \emph{qualitative} performance gap compared to all flagship models. Its Call Completeness (56.6\%) and Parameter Validity (58.6\%) fall roughly 30--40 percentage points below the flagship floor, indicating that the model systematically omits large segments of the required tool-call pipeline and frequently generates invalid parameters. Call Efficiency drops to 63.6\%---substantially below the flagship range of 97--100\%---reflecting a high proportion of redundant or erroneous calls. This gap is not merely quantitative: manual inspection reveals that Qwen 2.5-72B frequently fails to complete entire pipeline stages (e.g., omitting all camera track construction or skipping character animation assignment), rather than exhibiting the incremental omission patterns observed in weaker flagship models.

\subsubsection{Layer 2: Structural Integrity}

Table~\ref{tab:l2_results} reports the structural integrity of generated Level Sequences, evaluated programmatically on the final sequence state.

\begin{table}[h]
\centering
\begin{tabular}{@{}lccc@{}}
\toprule
\textbf{Model} & \textbf{TC}$\uparrow$ & \textbf{CamC}$\uparrow$ & \textbf{TempC}$\uparrow$ \\
\midrule
Claude Opus 4.6      & 100.0 &  96.4 &  99.5 \\
Claude Sonnet 4.6    &  99.6 &  89.5 &  98.6 \\
GPT-5.4              &  96.0 &  93.5 &  98.0 \\
Qwen 3.5 Plus        &  97.9 &  89.3 &  96.3 \\
Kimi K2.5            &  91.0 &  73.9 &  89.2 \\
GLM-5                &  92.4 &  77.3 &  95.8 \\
MiniMax M2.5         &  94.8 &  74.8 &  85.3 \\
Qwen 2.5-72B         &  50.9 &  66.2 &  50.1 \\
\bottomrule
\end{tabular}
\caption{Layer 2 evaluation results averaged over all test scenarios. All values in \%.}
\label{tab:l2_results}
\end{table}

Among the flagship models, Camera Coverage exhibits the widest performance gap: Claude Opus 4.6 achieves 96.4\%, while Kimi K2.5 (73.9\%), MiniMax M2.5 (74.8\%), and GLM-5 (77.3\%) fail to cover roughly a quarter of the effective sequence duration with active camera assignments, leaving extended periods with no camera active. Track Completeness correlates strongly with L1 Call Completeness---models that miss essential tool calls also produce incomplete track structures. Temporal Consistency further differentiates model quality: MiniMax M2.5 (85.3\%) and Kimi K2.5 (89.2\%) frequently produce animation overlaps and audio--facial misalignments, while the top-performing models maintain near-perfect temporal coherence ($\geq$98.0\%).

Qwen 2.5-72B's L2 results further confirm the severity of the middle-size model gap: Track Completeness drops to 50.9\% and Temporal Consistency to 50.1\%, indicating that roughly half of the expected character tracks are missing or temporally malformed. The generated sequences frequently lack entire camera cut tracks and exhibit severe asset binding failures, rendering the output unsuitable for downstream cinematic evaluation. Based on these results, we exclude Qwen 2.5-72B from subsequent L3 evaluation, as the generated Level Sequences do not produce renderable cutscenes of sufficient structural integrity for meaningful narrative and cinematic assessment. This exclusion itself constitutes a significant finding: cutscene generation---with its long-horizon multi-step tool calling, strict inter-tool dependencies, and stateful side effects---imposes a substantial capability threshold that current middle-size models do not meet.

\subsubsection{Layer 3: Narrative and Cinematic Quality}

Table~\ref{tab:l3_results} reports narrative and cinematic quality scores as assessed by a multimodal LLM judge (Gemini 3.1 Pro) on 25 sampled rendered videos per model.

\begin{table}[h]
\centering
\begin{tabular}{@{}lccccc@{}}
\toprule
\textbf{Model} & \textbf{SF}$\uparrow$ & \textbf{ChC}$\uparrow$ & \textbf{CQ}$\uparrow$ & \textbf{TmpCoh}$\uparrow$ & \textbf{Total}$\uparrow$ \\
\midrule
Claude Opus 4.6      &  10.8 &  14.1 &  13.2 &   12.1 &  50.2 \\
Claude Sonnet 4.6    &  10.4 &  11.7 &   9.8 &    9.8 &  41.7 \\
GPT-5.4              &  10.2 &  12.2 &  10.0 &   10.0 &  42.4 \\
Qwen 3.5 Plus        &   7.4 &   9.7 &   5.7 &    7.2 &  30.0 \\
Kimi K2.5            &   8.6 &   9.6 &   5.4 &    7.1 &  30.7 \\
GLM-5                &   8.0 &   8.4 &   5.7 &    6.8 &  28.9 \\
MiniMax M2.5         &   7.5 &   7.6 &   4.4 &    6.2 &  25.8 \\
\bottomrule
\end{tabular}
\caption{Layer 3 evaluation results (LLM-as-Judge). Each dimension is scored 0--25; Total is 0--100. Scores are averaged over 25 sampled videos per model (5 per scenario tier). Judge model: Gemini 3.1 Pro.}
\label{tab:l3_results}
\end{table}

Claude Opus 4.6 leads with a total score of 50.2, nearly double the lowest-scoring MiniMax M2.5 (25.8). The top three models (Opus 4.6, GPT-5.4, Sonnet 4.6) form a distinct upper cluster ($\geq$41.7), separated by a gap of over 10 points from the remaining four models ($\leq$30.7). Cinematographic Quality (CQ) is the most discriminative dimension, spanning an 8.8-point range (4.4--13.2): top-tier models produce coherent shot compositions and meaningful camera transitions, while lower-tier models frequently default to static wide shots with minimal camera variety. Script Fidelity (SF) shows the narrowest spread (7.4--10.8), indicating that basic narrative content delivery is a broadly shared capability---the primary differentiator lies in \emph{how} the narrative is visually realized rather than \emph{whether} script content is present. Note that absolute L3 scores reflect the judge model's conservative calibration, a known characteristic of LLM-as-Judge evaluations~\cite{zheng2023judging}; the relative rankings and inter-model gaps are the primary indicators of comparative quality.

Comprehensive per-scenario breakdowns across all evaluation layers and complexity scaling analysis are provided in Appendix~\ref{app:eval_details}.

\subsubsection{Aggregate Visualization}

\begin{figure}[h]
    \centering
    \includegraphics[width=0.75\textwidth]{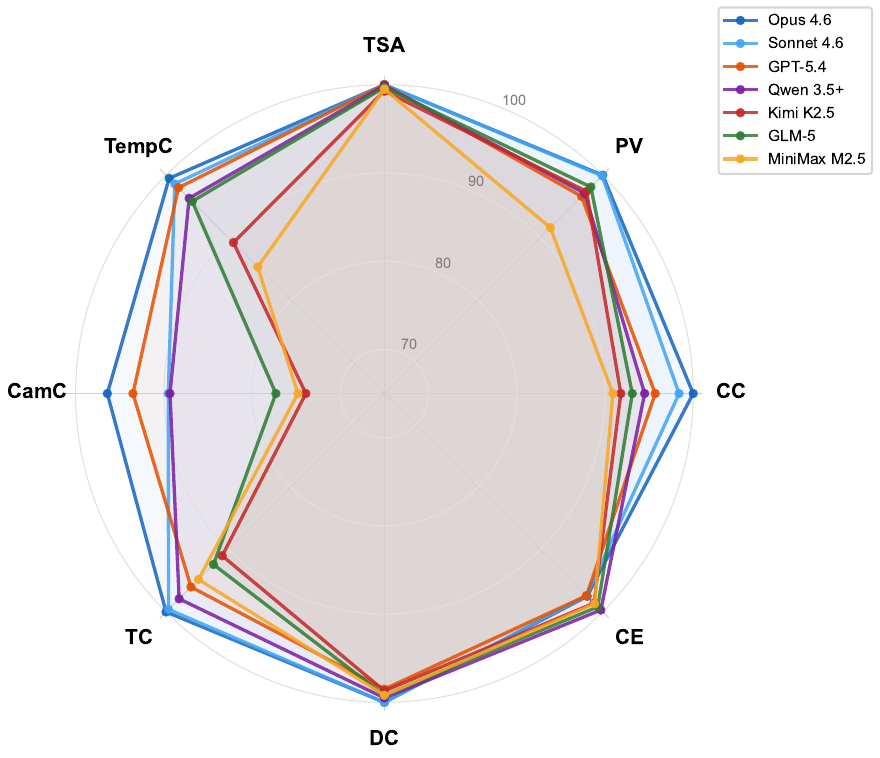}
    \caption{Radar chart comparing model performance across L1 and L2 metrics of CutsceneBench. Each axis represents one evaluation metric (range 65--100\%); values closer to the outer boundary indicate better performance. Camera Coverage (CamC) shows the widest inter-model spread.}
    \label{fig:eval_radar}
\end{figure}

\begin{figure}[h]
    \centering
    \includegraphics[width=0.8\textwidth]{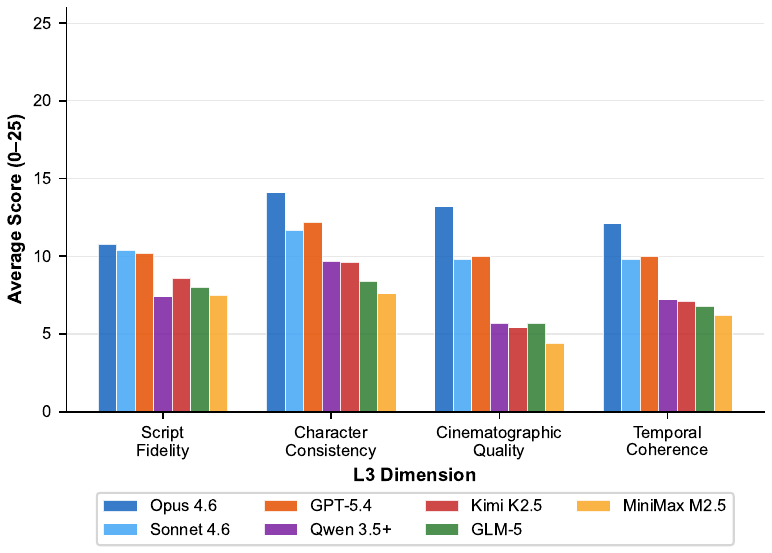}
    \caption{Layer 3 per-dimension scores (0--25) for all seven models. Cinematographic Quality (CQ) exhibits the widest inter-model gap; Script Fidelity (SF) is the most compressed dimension.}
    \label{fig:eval_l3_dims}
\end{figure}

Figure~\ref{fig:eval_radar} presents a radar chart visualization of aggregate model performance across all L1 and L2 metrics, while Figure~\ref{fig:eval_l3_dims} shows the per-dimension L3 scores on a separate scale.

\begin{figure}[h]
    \centering
    \includegraphics[width=0.9\textwidth]{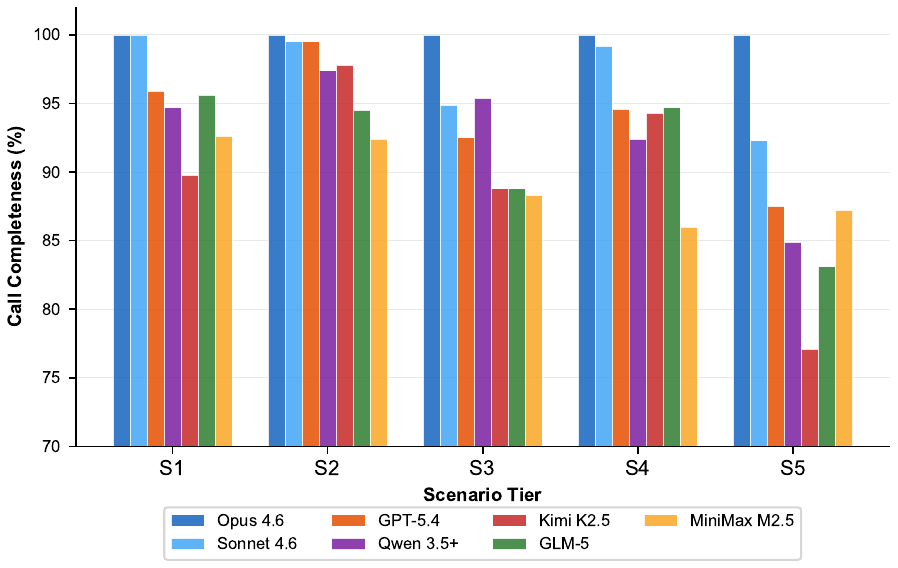}
    \caption{Per-scenario Call Completeness breakdown across CutsceneBench test scenarios (S1--S5). Performance generally decreases with task complexity, with the gap between top-performing and weaker models widening for complex scenarios.}
    \label{fig:eval_scenario}
\end{figure}

Figure~\ref{fig:eval_scenario} presents a per-scenario breakdown, illustrating how model performance varies with task complexity.

\subsection{Analysis}
\label{sec:eval_analysis}

We highlight several key findings from the evaluation results.

\noindent\textbf{Cross-Layer Correlation.} Models achieving high L1 Call Completeness consistently produce high L2 Track Completeness (Pearson $r > 0.95$), confirming that missing essential tool calls directly causes structural deficiencies in the output sequence. However, Temporal Consistency (L2) is not fully predicted by L1 metrics---a model can produce all required tool calls yet still generate overlapping animations or misaligned audio-facial tracks, indicating that \emph{parameter quality} matters beyond mere call presence. L3 scores broadly align with L1--L2 rankings but introduce additional separation: the top three models by L3 total (Opus 50.2, GPT-5.4 42.4, Sonnet 41.7) all achieve $\geq$95.7\% CC, while the bottom four ($\leq$30.7 L3 total) cluster below 94.5\% CC, confirming that technical correctness is a necessary but not sufficient condition for cinematic quality.

\noindent\textbf{Complexity Scaling.} L1 performance degrades monotonically with scenario complexity for all models except Claude Opus 4.6, which maintains perfect scores across all tiers. The degradation is most pronounced in Call Completeness, where lower-ranking models lose up to 12.7 percentage points between simple (S1) and complex (S5) scenarios. In contrast, L3 scores exhibit a \emph{non-monotonic} complexity pattern: several models peak on S3 (two-character emotional scenes) rather than S1, suggesting that narrative richness enables better cinematic expression even as technical difficulty increases. S4 (three-character scenes) is the most challenging tier across both L1 and L3, indicating that multi-character spatial reasoning remains a bottleneck for current LLMs.

\noindent\textbf{Failure Mode Analysis.} The hierarchical evaluation structure enables precise failure diagnosis. We observe four dominant failure patterns: (1)~\emph{omitted pipeline steps}, where models skip audio-to-face-expression or character-orientation calls, primarily affecting CC and TC; (2)~\emph{parameter hallucination}, where models fabricate character names or animation identifiers not present in the available asset library, primarily affecting PV; (3)~\emph{temporal misalignment}, where animation end times exceed the start of subsequent animations, primarily affecting TempC; and (4)~\emph{incomplete camera coverage}, where camera cut sections fail to span the full sequence duration, leaving segments with no active camera---this pattern is especially pronounced in the lower-tier models (CamC $\leq$77.3\%). Pattern~(1) is the most common across all non-optimal models, while pattern~(4) is the most discriminative at both L2 (CamC range: 22.5 pp) and L3 (CQ range: 8.8 points).

\noindent\textbf{Performance Stratification.} Consistent across all three layers, the results reveal a clear four-tier performance structure. Claude Opus 4.6 forms the \emph{top tier} with near-perfect L1--L2 scores and the highest L3 total (50.2). Claude Sonnet 4.6, GPT-5.4, and Qwen 3.5 Plus form a competitive \emph{upper-middle tier} with L3 totals of 30.0--42.4 and L2 CamC $\geq$89.3\%. Kimi K2.5, GLM-5, and MiniMax M2.5 constitute a \emph{lower-middle tier} where CC drops below 93\%, CamC below 78\%, TempC below 96\%, and L3 totals remain at 25.8--30.7. Qwen 2.5-72B occupies a distinct \emph{bottom tier}: with CC at 56.6\%, TC at 50.9\%, and TempC at 50.1\%, it falls far below the operational threshold for producing structurally viable cutscenes. The sharp performance cliff between the flagship lower-middle tier and the middle-size bottom tier underscores that cutscene generation---a task demanding long-horizon multi-step tool orchestration, strict dependency management, and precise temporal reasoning---imposes stringent requirements on underlying model capacity. The most discriminative metrics per layer among flagship models are CC (9.1 pp range) at L1, CamC (22.5 pp range) at L2, and CQ (8.8-point range) at L3---all relating to the completeness and quality of cinematic coverage, suggesting this is the central capability gap among current LLMs for cutscene generation.

\section{Conclusion}
\label{sec:conclusion}

We have presented Cutscene Agent, an LLM-driven framework that generates editable Unreal Engine Level Sequences from natural-language scripts, directly addressing the editability gap that separates current AI-generated cinematic content from professional production workflows. The framework contributes a Cutscene Toolkit built on the Model Context Protocol that provides composable, engine-agnostic tools for character management, asset querying, parameterized camera templates, and real-time scene perception; an agent system that combines priority-based prompt management, automatic state injection, category-aware history compression, and flexible subagent delegation to sustain coherent generation over long horizons of dozens of sequential tool calls; and CutsceneBench, a hierarchical evaluation benchmark that assesses tool-use correctness, sequence structural integrity, and narrative quality across scenarios of increasing complexity. A closed-loop visual reasoning mechanism---enabling the agent to perceive rendered frames and iteratively refine its creative decisions---further bridges the gap between structural validity and aesthetic quality.

Evaluation across eight LLMs reveals substantial variation in multi-step orchestration capability: while the strongest model achieves near-perfect scores on both tool-call correctness and sequence structure metrics, other models exhibit notable degradation on complex scenarios requiring long-horizon planning with strict dependency ordering. These results highlight that cutscene generation poses a challenging and differentiated benchmark for agentic LLM capabilities, one that existing tool-use evaluations do not cover.

\paragraph{Limitations and Future Directions.}
Several limitations point toward promising future work. The current system targets dialogue-driven cutscenes---expanding to action choreography, large crowd scenes, and dynamic environment interactions through additional gameplay track tools and motion planning is a natural next step. Generation quality is bounded by the available asset libraries, and the external TTS and facial animation services introduce pipeline latency; tighter integration of on-device generative models could alleviate both constraints. Cross-engine portability---adapting the MCP toolkit to Unity, Blender, or other DCC tools by reimplementing only the engine-specific layer---would broaden applicability without modifying the agent logic, while a human-in-the-loop co-creation workflow would leverage the inherently editable Level Sequence output for seamless handoff between AI generation and artist refinement. Beyond individual cutscenes, extending the framework to generate coherent sequences of cutscenes spanning entire narrative arcs---maintaining character state, visual continuity, and story progression across scenes---would bring the system closer to full-scale cinematic production.

\newpage
\phantomsection
\section*{Contributions}
\label{sec:contributions}

\noindent \textbf{Team Leader:} Qi Gan

\vspace{0.15cm}

\noindent \textbf{Project Leader:} Haozhou Pang

\vspace{0.15cm}

\noindent \textbf{Technical Implementation:} Lanshan He$^{*}$, Haozhou Pang$^{*}$, Qi Gan, Xin Shen, Ziwei Zhang, Yibo Liu, Gang Fang, Bo Liu, Kai Sheng, Shengfeng Zeng

\vspace{0.15cm}

\noindent \textbf{Artists \& Designers:} Chaofan Li, Zhen Hui,  Keer Zhou, Lan Zhou, Shujun Dai

\vspace{0.5cm}

\noindent {\small $^{*}$ Equal Contribution}

\newpage
\bibliographystyle{plain}
\bibliography{references}

\newpage
\appendix
\appendixtocheader

\section{Toolkit and MCP API Reference}
\label{app:api_reference}

This appendix serves as the comprehensive technical reference for the Cutscene Toolkit. We first describe two key architectural mechanisms---the \textbf{custom module system} for project-specific adaptation (\S\ref{app:custom_module}) and the \textbf{asset management architecture} with its data isolation and query subsystems (\S\ref{app:asset_arch})---followed by the complete MCP tool API specifications organized by the four functional modules described in Section~\ref{sec:toolkit}.

\subsection{Custom Module Mechanism}
\label{app:custom_module}

A cutscene toolkit intended for broad adoption must accommodate the significant variation in how different Unreal Engine projects organize their character blueprints, animation assets, and facial animation pipelines. Hard-coding any single project's loading logic into the toolkit would make it non-portable; exposing engine-internal loading details (e.g., blueprint class paths) to LLM agents would invite hallucination and misuse. The custom module mechanism solves both problems through a \textbf{registry-based loader abstraction} that cleanly separates the engine-facing asset loading layer from the agent-facing MCP tool interface.

\subsubsection{AssetLoaderRegistry}

The core of this mechanism is the \texttt{Asset\-Loader\-Registry}, a global singleton that provides two decorator-based extension points:

\begin{itemize}[nosep]
    \item \code{@register_loader(loader_type)} --- Registers a function that handles \textit{loading an asset into a Level Sequence} (e.g., spawning a character blueprint, adding an animation section). The function signature follows a typed contract defined by the \texttt{LoaderCallable} Protocol: \code{(asset_id: str, asset_record: AssetRecord, tool_args: dict | None) -> Any}, where \code{AssetRecord} is a Pydantic model containing \code{identifier}, \code{loader_type}, \code{asset_kind}, \code{source} (\texttt{"static"} or \texttt{"dynamic"}), \code{public_data}, and \code{private_data}. The explicit \code{tool_args} parameter replaces the former implicit \code{**kwargs}, providing transparent parameter passing from upstream MCP tools.
    \item \code{@register_receiver(importable_type, loader_type, asset_kind)} --- Registers a function that handles \textit{post-processing of dynamically imported assets} (e.g., converting a WAV file into an importable Sound Wave, or converting an NPZ facial curve file into a UE Animation Asset). The function signature follows the \texttt{ReceiverCallable} Protocol: \code{(context: ImportContext) -> ReceiverResult}. \texttt{ImportContext} is a Pydantic model encapsulating the full import context (\code{identifier}, \code{importable_type}, \code{loader_type}, \code{asset_kind}, \code{raw_file_abs}, \code{source_type}, \code{metadata}), while \texttt{ReceiverResult} contains \code{success}, \code{asset_path}, and \code{message}. The decorator additionally declares \code{loader_type} and \code{asset_kind}, enabling the registry to automatically map from import types to downstream loader and query categories.
\end{itemize}

When any MCP tool (e.g., \code{add_character}) is invoked, the toolkit looks up the \texttt{loader\_type} field from the matching \texttt{AssetRecord} and dispatches to the corresponding registered function. This indirection means the MCP tool signatures remain \textit{completely stable} across projects---only the backend loader implementation changes.

\subsubsection{Project Isolation Architecture}

The toolkit enforces a two-package directory structure:

\begin{center}
\small
\begin{tabular}{lll}
\toprule
\textbf{Package} & \textbf{Path} & \textbf{Role} \\
\midrule
\code{cutscene_provider} & \code{Content/Python/cutscene_provider/} & Core toolkit \\
\code{cutscene_provider_custom} & \code{Content/Python/cutscene_provider_custom/} & Project-specific extensions \\
\bottomrule
\end{tabular}
\end{center}

At plugin startup, \code{init_unreal.py} imports the \code{cutscene_provider_custom} package, which triggers all \code{@register_loader} and \code{@register_receiver} decorators in that package to execute, populating the global registry. The LLM agent is entirely unaware of the custom layer's existence---it interacts only with the stable MCP tool interface. A project adopting the toolkit need only: (1)~populate the asset spreadsheet with its own assets and the appropriate \texttt{loader} type strings in the \texttt{loader} column, and (2)~implement the corresponding loader functions in \code{cutscene_provider_custom/}.

\subsection{Asset Management Architecture}
\label{app:asset_arch}

The asset management system provides a unified interface through which LLM agents can discover and query all available assets---whether pre-authored or generated at runtime---while enforcing strict information isolation that prevents agents from accessing engine-internal data.

\subsubsection{Static Asset Spreadsheet Standard}

Pre-authored assets are stored in a structured Excel workbook (\texttt{.xlsx}) where each worksheet corresponds to one asset type (e.g., \texttt{Characters}, \code{Animation_Male}). The spreadsheet uses a \textbf{multi-row header schema} with the following structure:

\begin{center}
\small
\begin{tabular}{clp{8.5cm}}
\toprule
\textbf{Row} & \textbf{Name} & \textbf{Description} \\
\midrule
1 & Primary Category & One of four values: \texttt{identifier}, \texttt{loader}, \texttt{public data}, or \texttt{private data}. Controls data visibility and routing. Merged cells span all columns belonging to the same category. \\
2 & Field Name & The specific field key (e.g., \texttt{name}, \texttt{gender}, \code{blueprint_path}). \\
3 & Data Type & Type annotation for automatic conversion: \texttt{str}, \texttt{float}, \texttt{int}, or \texttt{bool}. \\
4 & Field Description & (Optional) Natural-language description used to auto-generate the query instruction guide returned by \code{get_query_instruction()}. \\
\midrule
5+ & Data Rows & Actual asset records. \\
\bottomrule
\end{tabular}
\end{center}

The following shows a simplified example of how an asset spreadsheet is structured (columns abbreviated):

\begin{center}
\footnotesize
\begin{tabular}{c|c|cc|cc}
\toprule
\multicolumn{1}{c|}{\textbf{identifier}} & \multicolumn{1}{c|}{\textbf{loader}} & \multicolumn{2}{c|}{\textbf{public data}} & \multicolumn{2}{c}{\textbf{private data}} \\
\code{asset_id} & \code{loader_type} & \texttt{name} & \texttt{gender} & \code{class_name} & \code{blueprint_path} \\
str & str & str & str & str & str \\
Unique ID & Loader key & Display name & male/female & \textit{(hidden)} & \textit{(hidden)} \\
\midrule
char\_01 & metahuman\_character & Alice & female & BP\_Alice & /Game/BP/... \\
char\_02 & metahuman\_character & Bob & male & BP\_Bob & /Game/BP/... \\
\bottomrule
\end{tabular}
\end{center}

Each \code{asset_type} parameter in the query tools corresponds directly to a worksheet name.

\subsubsection{Public/Private Data Isolation}
\label{app:data_isolation}

The four primary categories in Row~1 determine data routing through two distinct pathways:

\begin{itemize}[nosep]
    \item \textbf{Agent-facing path} (via \code{get_assets_for_llm()}): Only \texttt{identifier} and \texttt{public data} fields are included in MCP tool responses. The agent sees a sanitized record:
\begin{pythonapi}
{"identifier": "char_01", "info": {"name": "Alice", "gender": "female"}}
\end{pythonapi}
    \item \textbf{Engine-facing path} (via \code{invoke_loader()}): The registered loader function receives the \textit{complete} record as an \texttt{AssetRecord} Pydantic model including all \texttt{private data} fields:
\begin{pythonapi}
AssetRecord(
    identifier="char_01",
    loader_type="metahuman_character",
    asset_kind="Characters",
    source="static",
    public_data={"name": "Alice", "gender": "female"},
    private_data={"class_name": "BP_Alice",
                  "blueprint_path": "/Game/BP/..."})
\end{pythonapi}
\end{itemize}

This separation is a deliberate security and robustness measure: it prevents hallucination of engine-internal paths by the LLM, avoids exposing project-proprietary asset structures in the conversation context, and ensures that adding or renaming internal assets requires only spreadsheet updates---no agent prompt changes.

\subsubsection{Lightweight Query DSL}

The \code{query_assets()} tool accepts an optional \texttt{filters} dictionary whose keys are \texttt{public data} field names and whose values are filter expressions supporting three modes:

\begin{center}
\small
\begin{tabular}{lll}
\toprule
\textbf{Mode} & \textbf{Syntax} & \textbf{Semantics} \\
\midrule
Exact match & \texttt{"male"} & Case-insensitive string equality \\
Regex & \texttt{"/.*warrior.*/"}  & \texttt{re.search(pattern, val, IGNORECASE)} \\
Numeric comparison & \texttt{">5.0"}, \texttt{"<=10"}, \texttt{"=3"} & Cast to float, then compare \\
\bottomrule
\end{tabular}
\end{center}

The intended workflow follows a progressive-disclosure pattern: the agent first calls \code{get_queryable_asset_types()} to discover available types, then \code{get_query_instruction(type)} to retrieve the filterable field schema and syntax guide (auto-generated from the spreadsheet's Row~4 descriptions), and finally \texttt{query\_assets(type, filters=\{...\})} to execute the query. This three-step pattern prevents the agent from constructing malformed queries or filtering on nonexistent fields.

\subsubsection{Dynamic Asset Pipeline}

Assets generated at runtime---such as TTS audio clips or audio-driven facial animation curves---are ingested through the public \code{import_dynamic_asset} tool, which supports three data-source modes (\texttt{base64}, \texttt{file\_path}, \texttt{url}). The import lifecycle proceeds as follows:

\begin{enumerate}[nosep]
    \item The agent or an external service calls \code{import_dynamic_asset}, specifying \texttt{data\_type} (an importable type, e.g., \texttt{audio\_wav}), \texttt{data\_source} (the raw payload or reference), \texttt{source\_type} (\texttt{base64} / \texttt{file\_path} / \texttt{url}), and an optional \texttt{identifier\_hint}.
    \item The toolkit resolves the data source: base64 payloads are decoded in memory; file paths are copied into the project's \texttt{Raw/dynamic/} directory; URLs are downloaded asynchronously via a generator-based non-blocking mechanism (\texttt{yield}-based suspension) to avoid freezing the editor. A normalized identifier is assigned following a \texttt{\{prefix\}\_\{hint\}\_\{hex\_suffix\}} scheme.
    \item The raw file is dispatched to the appropriate \code{@register_receiver} handler (selected by \texttt{data\_type}) via an \texttt{ImportContext} object. The receiver performs format-specific processing (e.g., WAV$\rightarrow$Sound Wave import) and returns a \texttt{ReceiverResult} containing the resulting UE asset path.
    \item The record---now carrying explicit \texttt{importable\_type}, \texttt{loader\_type}, and \texttt{asset\_kind} fields---is persisted in a JSON registry (\code{dynamic_registry.json}). Once registered, dynamic assets become queryable through the same \code{query_assets()} interface, filtered by \texttt{asset\_kind}.
\end{enumerate}

Agents can discover available import types and their expected metadata before importing: \code{get_importable_asset_types()} returns all registered types with descriptions and supported file extensions, while \code{get_import_guide(data\_type)} provides recommended metadata fields and import behavior details.

The \texttt{Dynamic\-Asset\-Manager} applies the same sanitization as the static layer: \code{get_records_for_llm()} returns only \texttt{identifier} and user-specified \texttt{metadata}, stripping internal fields like \code{raw_path} and \code{asset_path}.

\subsection{Character \& Track Management Tools}

These tools allow the agent to spawn characters, control their placement, and manipulate their tracks within the Sequencer timeline. Design rationale involves providing atomic operations that map cleanly to standard 3D animation workflows while abstracting pipeline complexities (e.g., dynamic blending and bone alignment).

\begin{pythonapi}
def add_character(name: str, identifier: str, location: list[float] = (0.0, 0.0, 0.0)) -> str:

def orient_character_to_center(names: list[str]) -> str:

def add_character_animation(
    character_name: str, identifier: str, start_time: float
) -> dict:

def add_character_audio(
    character_name: str, identifier: str, start_time: float, 
    end_time: float, speech_text: str = ""
) -> dict:

def add_character_facial_animation(
    character_name: str, identifier: str, start_time: float, 
    gender: Literal["male", "female"] = "male" # to be removed
) -> dict:
\end{pythonapi}

\subsection{Asset Management \& Query Tools}

The asset query module bridges the agent's contextual knowledge with the concrete 3D resources loaded in the actual Unreal Engine project. The design rationale centers on dynamic schema-based querying, so that new digital assets can be added into the project without re-prompting or hard-coding their details into the agent.

\begin{pythonapi}
def get_queryable_asset_types() -> str:

def get_query_instruction(asset_type: str) -> str:

def query_assets(
    asset_type: str,
    filters: Optional[dict] = None, 
    include_generated: str = "auto"
) -> str:

def get_available_characters() -> str:

def get_available_animations(gender: str = "male") -> str:
\end{pythonapi}

\subsection{Dynamic Asset Import Tools}

These tools provide a progressive-discovery interface for dynamic asset ingestion. Agents can first query what types are importable and what metadata is recommended, then perform the actual import.

\begin{pythonapi}
def get_importable_asset_types() -> str:

def get_import_guide(data_type: str) -> str:

def import_dynamic_asset(
    data_type: str,
    data_source: str,
    source_type: str,       # "base64" | "file_path" | "url"
    file_extension: str,
    identifier_hint: str = "",
    metadata: dict | None = None
) -> str:
\end{pythonapi}

\subsection{Camera Management Tools}

The camera tools bridge the semantic intent of the camera agent with the rigorous spatial mathematics required to implement the camera template framework. The design rationale ensures the camera is driven primarily by abstract framing configurations and procedural offsets related to subject bones, preventing the model from having to guess raw coordinates.

\begin{pythonapi}
def add_camera(camera_name: str):

def set_active_camera(camera_name: str, start_time: float, end_time: float):

def get_available_camera_templates():

def apply_camera_template(
    camera_name: str,
    position_template: str,
    position_args: dict,
    movement_template: Optional[str] = None,
    movement_args: Optional[dict] = None,
    *,
    start_time: float,
    duration: float,
) -> str:
\end{pythonapi}

\subsection{Scene Perception \& Interaction Tools}

These tools serve as the ``eyes'' of the multi-agent system. The core design rationale is to supply explicit semantic feedback regarding spatial, temporal, and metadata states across tracks and bindings, avoiding reliance on blind operation.

\begin{pythonapi}
def update_sequence_metadata(new_block: dict) -> str:

def get_sequence_content() -> str:

def clear_sequence() -> str:

def set_current_sequence_time(time: float) -> str:

def move_view(
    forward: float = 0.0, horizontal: float = 0.0, vertical: float = 0.0, 
    yaw: float = 0.0, pitch: float = 0.0
) -> str:

def undo_move_view() -> str:

def take_editor_screenshot(resolution: list[int] = (1280, 720)) -> MCPImage:

def take_camera_screenshot(camera_name: str, resolution: list[int] = (1280, 720)) -> MCPImage:
\end{pythonapi}

\subsection{External Service Tools}

These tools connect external multimodal foundation models directly into the pipeline, typically mediated through Python backends independent of Unreal Engine. The rationale is to handle intensive computations like TTS and facial expression inference off-engine.

\begin{pythonapi}
def get_available_tone(character_name: str = "") -> list:

async def tts_function_tool(
    identifier: str, text: str, gender: str = "male", 
    tone: str = "", emotion: str = "normal"
) -> str:

async def audio_to_face_expression_tool(
    identifier: str, audio_identifier: str, emotion: str = "neutral"
) -> str:

def video_understanding_tool(video_identifier: str, task_description: str) -> str:
\end{pythonapi}


\section{Agent System Specifications}
\label{app:agent_specs}

This appendix provides the complete specifications of all agents in the multi-agent generation system, including prompt assembly, tool access scopes, and architectural designs for both implemented and planned components. It complements the overview in Section~\ref{sec:multiagent}.

\begin{figure}[htbp]
    \centering
    \includegraphics[width=\textwidth]{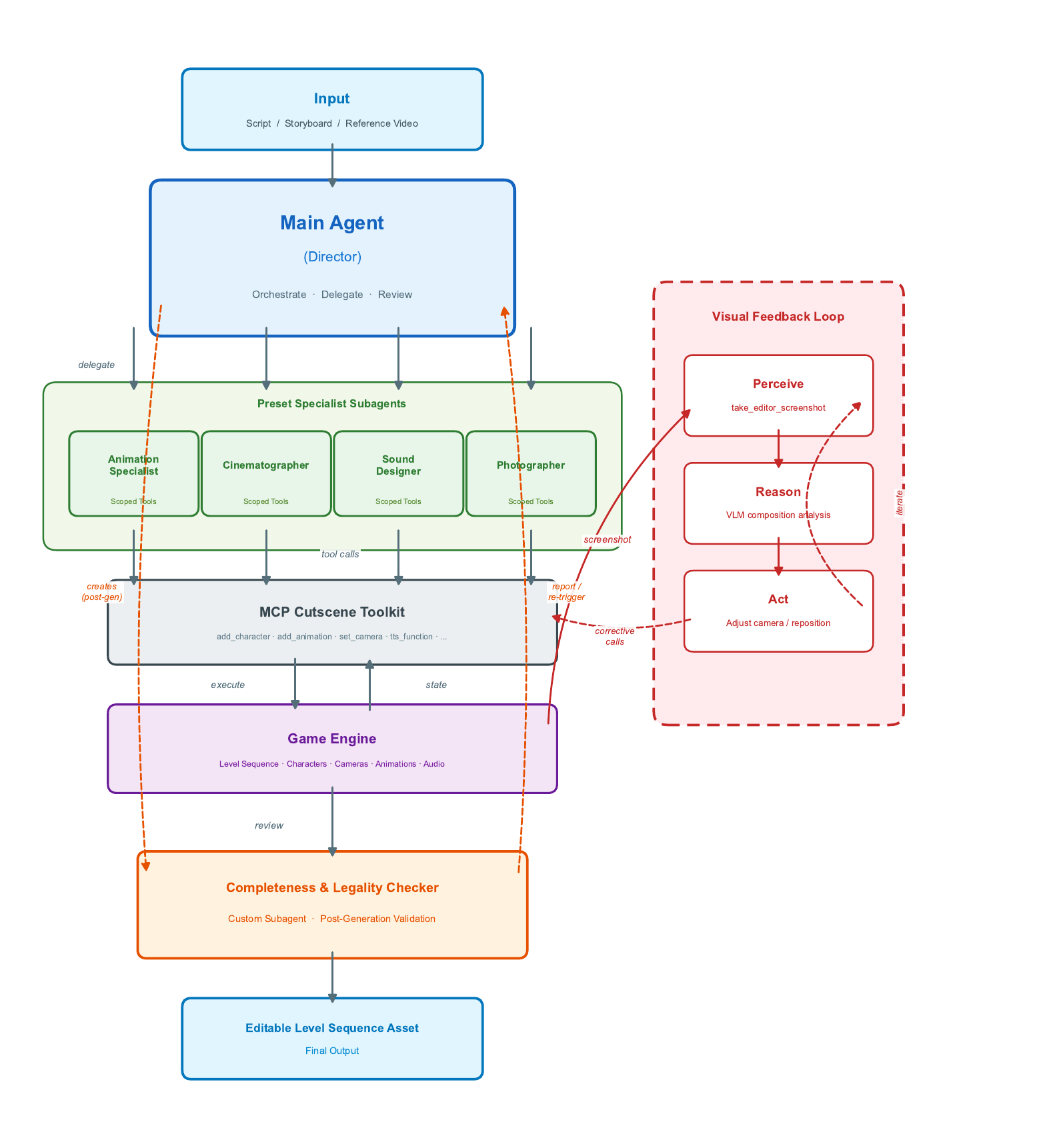}
    \caption{Overview of the Agent System architecture. The main agent (Director) orchestrates preset specialist subagents for domain-specific tasks and interacts with the game engine via the MCP Cutscene Toolkit. A visual feedback loop enables perception-driven refinement through multimodal LLMs. After generation, a dynamically created custom subagent performs completeness and legality validation.}
    \label{fig:architecture}
\end{figure}

\subsection{Director Agent Configuration}
\label{app:director}

The Director Agent serves as the top-level orchestrator, responsible for interpreting input scripts, planning shot structure, and delegating specialized subtasks to subagents. Its system prompt is dynamically assembled by the \texttt{PromptManager} from composable elements ranked by priority.

\subsubsection{Prompt Assembly Pipeline}

The prompt is constructed from a hierarchy of \texttt{PromptElement} objects. Each element carries a numeric priority; the \texttt{PromptManager} greedily selects elements in descending priority order until the token budget is exhausted, then renders them in category order (instructions first, context blocks second). Table~\ref{tab:prompt_priority} shows the full priority hierarchy.

\begin{table}[h]
\centering
\small
\begin{tabular}{clp{7.5cm}}
\toprule
\textbf{Priority} & \textbf{Category} & \textbf{Content} \\
\midrule
1000 & Identity        & Model awareness declaration and expert role definition \\
1000 & Safety          & Content policy compliance guardrails \\
1000 & Director Role   & Core directive: orchestrate cutscene creation and delegate to specialists \\
900  & Tool Usage      & Sequential calling discipline, dependency ordering, prerequisite enforcement \\
800  & Cutscene Rules  & Six domain-specific knowledge blocks (see below) \\
750  & Delegation      & Subagent invocation guidelines and return format \\
\bottomrule
\end{tabular}
\caption{Prompt element priority hierarchy for the Director Agent. Under context-window pressure, lower-priority elements are pruned first.}
\label{tab:prompt_priority}
\end{table}

\subsubsection{Domain Knowledge Blocks}

Six \texttt{ContextBlock} elements (all at priority 800) encode the Director's domain expertise, each wrapped in XML tags for structured parsing:

\begin{enumerate}[nosep]
    \item \textbf{CutsceneCreation} --- Canonical workflow ordering: add characters $\rightarrow$ generate audio and facial expressions $\rightarrow$ add body animations $\rightarrow$ add cameras and apply templates. Instructs the agent to consult the auto-injected \texttt{current\_cutscene\_content} for incremental operations.
    \item \textbf{Actor Rules} --- Character spatial placement conventions with concrete coordinate examples for two-person ($A$ at $(-60, 0, 0)$, $B$ at $(60, 0, 0)$) and three-person triangle formations ($A$, $B$, $C$ at prescribed offsets). Spacing heuristics: $\sim$70\,cm side-by-side, $\sim$40\,cm front-to-back.
    \item \textbf{Audio Rules} --- Tone consistency requirements (same character uses the same voice tone throughout a scene). Dialogue parsing guidelines: strip narration, identify emotional beats, compute \texttt{end\_time = start\_time + audio\_duration}.
    \item \textbf{Animation Rules} --- Animation tag taxonomy (\texttt{timing}: Speak/Gap/Solo; \texttt{expressiveness}: Light/Medium/Heavy; \texttt{mood}: emotional tone; \texttt{duration}). Duration-matching requirement: animation length must cover dialogue duration. No temporal overlap on the same character.
    \item \textbf{Camera Rules} --- Template-based camera positioning via \texttt{apply\_camera\_template}. Dynamic movement composition (Dolly push/pull, Orbit arc). OTS variant selection guidelines (\texttt{near}/\texttt{mid}/\texttt{high}). Encourages shot variety for visual richness.
    \item \textbf{Tool Usage} --- Sequential tool calling discipline, dependency ordering enforcement, and create-before-use prerequisites.
\end{enumerate}

\subsubsection{Tool Access Scope}

The Director Agent has access to the full MCP tool surface (30+ tools across all four functional modules described in Section~\ref{sec:toolkit}), with the following exceptions filtered out to prevent accidental misuse:

\begin{itemize}[nosep]
    \item \texttt{clear\_sequence} --- Prevents accidental deletion of in-progress work.
    \item \texttt{save\_sequence\_as} --- Prevents unintended overwrites of named assets.
    \item Screenshot tools --- Reserved for the visual feedback loop (Section~\ref{sec:visual_feedback}).
\end{itemize}

\noindent In addition to MCP tools, the Director has access to a local \texttt{run\_subagent} function tool for delegation, with the following interface:

\begin{pythonapi}
def run_subagent(
    template_name: str,     # Preset template or "custom"
    task: str,              # Natural-language task description
    context: str = "",      # Additional context for the subagent
    custom_instructions: str = "",   # (custom mode only)
    custom_tool_scope: list[str] = [] # (custom mode only)
) -> dict:  
    # Returns: {status, template_name, tool_calls_count,
    #           tool_calls[], result_summary, turns_used}
\end{pythonapi}

\subsection{Preset Specialist Subagents}
\label{app:specialists}

Each preset subagent is defined by a \texttt{SubAgentTemplate} comprising: (1)~a tailored system prompt, (2)~a tool whitelist enforced via MCP server-side filtering, and (3)~a maximum turn budget. All subagents receive the current Level Sequence state upon creation and operate with independent context windows. Table~\ref{tab:subagent_summary} provides an overview.

\begin{table}[h]
\centering
\small
\begin{tabular}{lp{4.8cm}cl}
\toprule
\textbf{Subagent} & \textbf{Responsibility} & \textbf{Tools} & \textbf{Max Turns} \\
\midrule
Scene Specialist       & Character spawning, spatial layout, orientation     & 4 & 20 \\
Animation Specialist   & Body animation selection, timing, emotional matching & 3 & 30 \\
Cinematographer        & Camera placement, shot composition, dynamic movement & 4 & 25 \\
Sound Designer         & TTS generation, audio attachment, facial animation   & 5 & 30 \\
Photographer           & Visual composition refinement via multimodal feedback & 3 & 10 \\
\bottomrule
\end{tabular}
\caption{Summary of preset specialist subagents. Tool counts exclude query/perception tools shared across agents.}
\label{tab:subagent_summary}
\end{table}

\subsubsection{Scene Specialist}

Responsible for adding characters to the Level Sequence and arranging their spatial positions and orientations according to stage directions.

\paragraph{System prompt (excerpt).}
\textit{You are a professional scene construction assistant responsible for adding and managing characters in Unreal Engine according to the director's instructions. Your duties: (1)~add characters to the scene at specified positions; (2)~arrange reasonable spatial layouts; (3)~adjust character orientations to face the scene center.}

\paragraph{Tool whitelist.}
\begin{center}
\small
\begin{tabular}{ll}
\toprule
\textbf{Tool} & \textbf{Purpose} \\
\midrule
\texttt{add\_character} & Spawn a character at a given position \\
\texttt{orient\_character\_to\_center} & Rotate characters to face the group centroid \\
\texttt{get\_available\_characters} & Query the character asset catalog \\
\bottomrule
\end{tabular}
\end{center}

\paragraph{Placement conventions.}
The system prompt encodes two standard formation templates:
\begin{itemize}[nosep]
    \item \textit{Two-person dialogue}: $A$ at $(-60, 0, 0)$, $B$ at $(60, 0, 0)$; call \texttt{orient\_character\_to\_center([A, B])}.
    \item \textit{Three-person triangle}: $A$ at $(-60, 0, 0)$, $B$ at $(30, -52, 0)$, $C$ at $(30, 52, 0)$; orient all three.
\end{itemize}
These are defaults; the director may override positions via the \texttt{task} description.

\subsubsection{Animation Specialist}

Selects and assigns body animations to characters based on dialogue content, emotional context, and temporal constraints.

\paragraph{System prompt (excerpt).}
\textit{You are a professional animation choreography assistant. Select and add appropriate body animations for characters, ensuring that animation durations match dialogue durations and that the emotional tone of each animation aligns with the script's intent.}

\paragraph{Tool whitelist.}
\begin{center}
\small
\begin{tabular}{ll}
\toprule
\textbf{Tool} & \textbf{Purpose} \\
\midrule
\texttt{add\_character\_animation} & Assign a body animation to a character at a given time \\
\texttt{get\_available\_animations} & Query the animation catalog with tag filters \\
\bottomrule
\end{tabular}
\end{center}

\paragraph{Animation selection taxonomy.}
The agent selects animations based on a structured tag system (see the asset query interface in Section~\ref{sec:toolkit}):
\begin{center}
\small
\begin{tabular}{lp{8.5cm}}
\toprule
\textbf{Tag} & \textbf{Semantics} \\
\midrule
\texttt{timing} & When to use: \texttt{Speak} (accompanies dialogue), \texttt{Gap} (between lines), \texttt{Solo} (standalone performance) \\
\texttt{expressiveness} & Gesture magnitude: \texttt{Light} (subtle), \texttt{Medium} (moderate), \texttt{Heavy} (emphatic) \\
\texttt{mood} & Emotional tone (e.g., \texttt{happy}, \texttt{angry}, \texttt{sad}); \texttt{null} if mood-neutral \\
\texttt{duration} & Animation clip length in seconds \\
\bottomrule
\end{tabular}
\end{center}

\noindent Key behavioral rules: (1)~animation duration must cover the corresponding dialogue segment; (2)~no temporal overlap is permitted on the same character's track; (3)~if the script lacks explicit stage directions, the agent infers appropriate animations from dialogue sentiment.

\subsubsection{Cinematographer}

Manages camera creation, shot composition via the parametric template system, and dynamic camera movement.

\paragraph{System prompt (excerpt).}
\textit{You are an expert cinematographer. Place cameras using the template-based positioning system, compose shots following established cinematographic conventions (shot--reverse-shot for dialogue, establishing shots for scene openings), and apply dynamic movement templates (Dolly, Orbit) for visual interest.}

\paragraph{Tool whitelist.}
\begin{center}
\small
\begin{tabular}{ll}
\toprule
\textbf{Tool} & \textbf{Purpose} \\
\midrule
\texttt{add\_camera} & Create a new camera actor \\
\texttt{set\_active\_camera} & Assign a camera to a time range on the camera cut track \\
\texttt{apply\_camera\_template} & Position and animate a camera using templates \\
\texttt{get\_available\_camera\_templates} & List available position/movement templates \\
\bottomrule
\end{tabular}
\end{center}

\paragraph{Cinematographic guidelines.}
The system prompt encodes the following heuristics: (1)~use OTS (over-the-shoulder) templates for dialogue coverage with alternating \texttt{from\_actor}/\texttt{to\_actor} for shot--reverse-shot patterns; (2)~open scenes with \texttt{Establishing} templates to reveal spatial context; (3)~apply \texttt{Dolly} with $\rho < 1$ for dramatic emphasis, \texttt{Orbit} for visual dynamism during monologues; (4)~vary shot types across consecutive cuts to avoid visual monotony. The complete template specifications are given in Appendix~\ref{app:camera_specs}.

\subsubsection{Sound Designer}

Handles the complete audio-to-facial-animation pipeline: TTS voice generation, audio track assignment, and audio-driven facial expression synthesis. This subagent manages the most complex dependency chain in the system.

\paragraph{System prompt (excerpt).}
\textit{You are a professional sound design and voice performance assistant. Generate character voices via TTS with consistent tone assignment per character, attach audio tracks to the timeline, and drive facial animation from audio signals. Maintain strict pipeline ordering: TTS generation $\rightarrow$ audio track attachment $\rightarrow$ facial expression synthesis $\rightarrow$ facial animation attachment.}

\paragraph{Tool whitelist.}
\begin{center}
\small
\begin{tabular}{ll}
\toprule
\textbf{Tool} & \textbf{Purpose} \\
\midrule
\texttt{tts\_function\_tool} & Generate speech audio from text with voice/emotion control \\
\texttt{add\_character\_audio} & Attach an audio asset to a character's audio track \\
\texttt{audio\_to\_face\_expression\_tool} & Generate facial animation curves from audio \\
\texttt{add\_character\_facial\_animation} & Attach facial animation to a character's face track \\
\texttt{get\_available\_tone} & Query available voice tones per character \\
\bottomrule
\end{tabular}
\end{center}

\paragraph{Pipeline dependency chain.}
The Sound Designer must respect a strict ordering due to data dependencies between external services and the engine:

\begin{enumerate}[nosep]
    \item \textbf{TTS generation}: Call \texttt{tts\_function\_tool} to synthesize speech $\rightarrow$ returns an \texttt{audio\_identifier} and \texttt{duration}.
    \item \textbf{Audio attachment}: Call \texttt{add\_character\_audio} with the audio identifier and computed \texttt{end\_time = start\_time + duration}.
    \item \textbf{Facial expression synthesis}: Call \texttt{audio\_to\_face\_expression\_tool} with the audio identifier $\rightarrow$ returns a \texttt{face\_identifier}.
    \item \textbf{Facial animation attachment}: Call \texttt{add\_character\_facial\_animation} with the face identifier.
\end{enumerate}

\noindent A critical behavioral constraint is \textbf{tone consistency}: the system prompt requires the agent to call \texttt{get\_available\_tone} at the start of each session, assign a fixed tone to each character, and reuse that assignment across all dialogue lines. Emotional variation is achieved via the \texttt{emotion} parameter rather than tone switching.

\subsubsection{Photographer (Visual Feedback Specialist)}
\label{app:photographer}

Unlike the template-based subagents above, the Photographer operates as a \textbf{standalone agent} with its own execution loop and multimodal perception pipeline. It implements the Perceive--Reason--Act cycle described in Section~\ref{sec:visual_feedback}.

\paragraph{System prompt (excerpt).}
\textit{You are an expert virtual photographer agent who controls the viewport camera. Analyze the current visual composition from screenshots, identify framing issues (occlusion, poor headroom, rule-of-thirds violations), and issue corrective camera adjustments. Continue refining until the composition meets quality criteria.}

\paragraph{Tool whitelist.}
\begin{center}
\small
\begin{tabular}{ll}
\toprule
\textbf{Tool} & \textbf{Purpose} \\
\midrule
\texttt{move\_view} & Relative camera translation and rotation (game-like controls) \\
\texttt{undo\_move\_view} & Revert the last camera adjustment \\
\texttt{take\_editor\_screenshot} & Capture the current viewport as an image \\
\bottomrule
\end{tabular}
\end{center}

\paragraph{Architectural differences.}
The Photographer differs from template-based subagents in several key respects:

\begin{itemize}[nosep]
    \item \textbf{Multimodal context}: After every tool execution step, the agent automatically captures a screenshot, resizes it to a normalized resolution (default $640 \times 480$), and injects it into the conversation as a vision message. A sliding window retains the $k = 3$ most recent images to provide temporal context.
    \item \textbf{Iterative refinement}: The agent runs in a tight loop (max 10 turns) where each iteration consists of: analyze current image $\rightarrow$ identify composition issues $\rightarrow$ execute corrective \texttt{move\_view} $\rightarrow$ capture new screenshot $\rightarrow$ re-evaluate. The loop terminates when the agent judges the composition satisfactory.
    \item \textbf{Independent execution}: The Photographer maintains its own \texttt{Runner} instance separate from the Director's execution loop, allowing it to be invoked both as a subagent and as an interactive standalone tool.
\end{itemize}

\subsection{Custom Agent Mode}
\label{app:custom_mode}

Beyond the preset specialist templates, the Director Agent can dynamically construct \textbf{custom subagents} at runtime to handle ad-hoc tasks not covered by existing presets.

\subsubsection{Runtime Subagent Construction}

The Director invokes the custom mode by calling \texttt{run\_subagent} with \texttt{template\_name="custom"} and two additional parameters:

\begin{itemize}[nosep]
    \item \texttt{custom\_instructions}: A free-form natural-language string that becomes the custom subagent's system prompt. The Director writes these instructions based on the specific task requirements, encoding behavioral guidelines, success criteria, and any domain knowledge relevant to the subtask.
    \item \texttt{custom\_tool\_scope}: A list of MCP tool names that the custom subagent is permitted to call. The \texttt{SubAgentRunner} enforces this whitelist via server-side tool filtering, ensuring the custom agent cannot access tools outside its designated scope.
\end{itemize}

\noindent The custom subagent's full lifecycle---context initialization (injecting current cutscene state), execution (up to a configurable turn limit), result collection, and resource cleanup (MCP server disconnection)---is managed entirely by the \texttt{SubAgentRunner} without human intervention. This mechanism allows the agent system to \textit{extend its own capability surface} at runtime.

\subsubsection{Use Cases}

Custom agents are particularly effective for:

\begin{itemize}[nosep]
    \item \textbf{Targeted review tasks}: The Director can spawn a review agent with read-only tool access (e.g., only \texttt{get\_sequence\_content} and screenshot tools) to audit the current cutscene for specific quality criteria before finalizing.
    \item \textbf{Specialized corrections}: When the Director identifies a specific issue (e.g., audio timing misalignment across multiple tracks), it can create a focused correction agent with only the relevant mutation tools and highly specific instructions.
    \item \textbf{Project-specific workflows}: Teams can define reusable custom agent configurations for recurring tasks unique to their production pipeline (e.g., a ``lighting review agent'' with custom evaluation heuristics).
\end{itemize}

\subsection{Vision-Language Agent Architectures}
\label{app:vla}

The Photographer agent (Section~\ref{app:photographer}) demonstrates that multimodal perception significantly enhances generation quality. This section describes a broader architectural vision for integrating \textbf{Vision-Language Agents (VLAs)} into the cutscene pipeline as specialized perception and evaluation modules. These designs represent planned extensions; architectural interfaces are specified but implementations are deferred to future work.

\subsubsection{General VLA Integration Pattern}

All VLA agents share a common three-stage architecture that interfaces with the existing \texttt{run\_subagent} framework:

\begin{enumerate}[nosep]
    \item \textbf{Perceive}: The VLA receives one or more screenshots captured via \texttt{take\_editor\_screenshot} or \texttt{take\_camera\_screenshot}, along with a structured task description and relevant cutscene metadata (character names, intended shot type, script excerpt).
    \item \textbf{Analyze}: A vision-language model (e.g., multimodal LLMs or dedicated vision-language architectures) processes the visual input against task-specific evaluation criteria, producing a structured assessment.
    \item \textbf{Report/Act}: The VLA returns either (a)~a \textit{diagnostic report}---structured JSON with quality scores, identified issues, and corrective suggestions---for the Director to act upon, or (b)~\textit{direct tool calls} to autonomously apply corrections.
\end{enumerate}

\noindent The key design principle is \textbf{interface uniformity}: VLA agents plug into the same \texttt{run\_subagent} dispatch mechanism as text-only specialists, differing only in their internal use of vision-language models and their access to screenshot tools. This ensures that adding new VLA specialists requires no changes to the orchestration framework.

\subsubsection{Planned VLA Specialist Designs}

We outline four specialist VLA agents targeting distinct quality dimensions:

\paragraph{Scene Layout Validator.}
Verifies that character spatial arrangements match the script's stage directions.
\begin{itemize}[nosep]
    \item \textit{Input}: Editor screenshot + script excerpt with stage directions (e.g., ``Alice stands to Bob's left'').
    \item \textit{Analysis}: Identifies each character in the rendered image, estimates their relative positions, and cross-references against the script's spatial specifications.
    \item \textit{Output}: Per-character spatial correctness scores and specific violation descriptions (e.g., ``Alice is positioned to Bob's right instead of left'').
    \item \textit{Action mode}: Can issue \texttt{add\_character} repositioning calls to correct violations.
\end{itemize}

\paragraph{Composition Quality Assessor.}
Evaluates camera shot quality against established cinematographic principles.
\begin{itemize}[nosep]
    \item \textit{Input}: Camera screenshot + intended shot type (e.g., OTS, close-up) and subject character.
    \item \textit{Analysis}: Evaluates rule-of-thirds adherence, headroom, lead room, depth separation, and subject prominence. Cross-references the intended shot type against the actual framing.
    \item \textit{Output}: Multi-dimensional quality score vector (composition, framing, depth) with per-dimension diagnostic and suggested template parameter adjustments.
\end{itemize}

\paragraph{Continuity Checker.}
Detects visual inconsistencies across consecutive shots that would break narrative immersion.
\begin{itemize}[nosep]
    \item \textit{Input}: Ordered sequence of screenshots spanning a camera cut transition.
    \item \textit{Analysis}: Checks for character appearance consistency (clothing, pose continuity), spatial coherence (180-degree rule compliance), lighting consistency, and prop placement across the cut.
    \item \textit{Output}: Per-cut continuity violation report with severity ratings and specific frame-pair comparisons.
\end{itemize}

\paragraph{Reference Style Matching Agent.}
Aligns the visual style of generated cutscenes with reference footage.
\begin{itemize}[nosep]
    \item \textit{Input}: Reference video frames (extracted via \texttt{video\_understanding\_tool}) + current cutscene screenshots.
    \item \textit{Analysis}: Compares shot composition patterns, camera movement dynamics, pacing rhythms, and framing preferences between reference and generated content.
    \item \textit{Output}: Style deviation report with actionable parameter adjustments (e.g., ``reference uses tighter framing---reduce OTS \texttt{distance\_back} by 30\%'').
\end{itemize}

\subsection{IDE Integration Variant}
The agent system naturally extends to interactive development environments. We implement a variant that maps the agent system's subagent primitives onto the native agent customization framework of VS Code Copilot (\texttt{.agent.md} files), enabling developers to invoke specialist subagents directly from their IDE. In this configuration, each preset subagent is declared as an \texttt{.agent.md} file specifying: a YAML frontmatter with tool whitelists, sub-agent references, and invocability flags; and a Markdown body containing the system prompt. Domain knowledge rules are modularized into separate reference files (e.g., \texttt{rules-actor.md}, \texttt{rules-camera.md}) included by the skill system, allowing selective loading based on task context. This integration demonstrates the portability of the agent specifications beyond the standalone orchestration framework.


\section{Camera Template Specifications}
\label{app:camera_specs}

This appendix details the mathematical formulations for computing camera positions and trajectories from character skeletal data.
All computations operate in Unreal Engine's left-handed, $Z$-up world coordinate system with positions in centimeters and angles in degrees.

\paragraph{Notation.}
Let $\mathbf{p}_A$ denote the world-space root location of actor~$A$, and $\mathbf{b}_{A}^{(\text{bone})}$ the world-space translation of a named skeletal bone on actor~$A$ (obtained via \texttt{Get\-Socket\-Transform} in world space).
We write $\hat{\mathbf{v}} = \mathbf{v} / \lVert\mathbf{v}\rVert$ for unit normalization, $\hat{\mathbf{z}} = (0,0,1)$ for the world up vector, and $\textsc{LookAt}(\mathbf{o}, \mathbf{t})$ for the engine rotation that orients a camera at origin~$\mathbf{o}$ to face target~$\mathbf{t}$.
$\textsc{Rot}(\mathbf{v}, \theta, \hat{\mathbf{a}})$ denotes the Rodrigues rotation of vector~$\mathbf{v}$ by angle~$\theta$ around axis~$\hat{\mathbf{a}}$.

\subsection{Position Templates}
\label{app:position_templates}

The toolkit provides six parameterized position templates.
Table~\ref{tab:position_summary} summarizes the key characteristics; the subsections that follow give the complete parameter schema and geometric derivation for each.

\begin{table}[h]
\centering
\small
\begin{tabular}{llcl}
\toprule
\textbf{Template} & \textbf{Typical Use} & \textbf{Actors} & \textbf{Primary Control} \\
\midrule
OTS              & Dialogue coverage     & $\geq$2 & \texttt{variant} preset \\
POV              & Subjective moments    & $\geq$2 & head offsets \\
OnAxis           & Direct confrontation  & $\geq$2 & head midpoint \\
SideProfile      & Reaction shots        & 1 & side, distance \\
Establishing     & Scene openers         & $\geq$2 & side, distance, height \\
GenericFocus     & Flexible framing      & 1 & spherical coords \\
\bottomrule
\end{tabular}
\caption{Position template summary.}
\label{tab:position_summary}
\end{table}

\subsubsection{Over-the-Shoulder (OTS)}

The OTS template positions the camera behind one character's shoulder, focusing on the other character.
Three \texttt{variant} presets provide typical framing for different dramatic contexts:

\begin{center}
\small
\begin{tabular}{lccc}
\toprule
\textbf{Variant} & $h_{\text{off}}$ & $d_{\text{side}}$ & $d_{\text{back}}$ \\
\midrule
\texttt{near}  & 40  & 100 & 140 \\
\texttt{mid}   & 50  & 230 & 200 \\
\texttt{high}  & 120 & 200 & 300 \\
\bottomrule
\end{tabular}
\end{center}

\paragraph{Parameters.}
\begin{center}
\footnotesize
\begin{tabular}{llll}
\toprule
\textbf{Parameter} & \textbf{Type} & \textbf{Default} & \textbf{Description} \\
\midrule
\code{from_actor_name} & str & (required) & Actor whose shoulder appears in foreground \\
\code{to_actor_name}   & str & (required) & Actor in focus \\
\texttt{variant}           & \texttt{near$|$mid$|$high} & \texttt{mid} & Preset framing variant \\
\code{shoulder_bone_name}     & str   & \texttt{head} & Reference bone for shoulder height \\
\code{shoulder_height_offset} & float & \textit{auto} & Vertical offset from bone (overrides preset) \\
\code{shoulder_side_offset}   & float & \textit{auto} & Lateral offset (overrides preset) \\
\code{distance_back}          & float & \textit{auto} & Backward distance (overrides preset) \\
\bottomrule
\end{tabular}
\end{center}

\paragraph{Geometric computation.}
Let $\mathbf{p}_F, \mathbf{p}_T$ be the root locations of the \emph{from} and \emph{to} actors, and $h_F^{(s)}, h_T^{(s)}$ the world-space $z$-coordinates of their shoulder bones. Compute the forward and right basis vectors:
\begin{align}
    \hat{\mathbf{f}} &= \widehat{\mathbf{p}_T - \mathbf{p}_F}, &
    \hat{\mathbf{r}} &= \widehat{\hat{\mathbf{z}} \times \hat{\mathbf{f}}}
\end{align}
The camera position is:
\begin{equation}
    \mathbf{p}_{\text{cam}} = \mathbf{p}_F
        - \hat{\mathbf{f}} \cdot d_{\text{back}}
        + \hat{\mathbf{r}} \cdot d_{\text{side}}
        + (0, 0,\; h_F^{(s)} + h_{\text{off}})
\end{equation}
The look-at target is the midpoint of the two actors' shoulder positions:
\begin{equation}
    \mathbf{t}_{\text{look}} = \tfrac{1}{2}\bigl((\mathbf{p}_F + (0,0,h_F^{(s)})) + (\mathbf{p}_T + (0,0,h_T^{(s)}))\bigr)
\end{equation}

\subsubsection{Point-of-View (POV)}

The POV template approximates a character's viewpoint, positioned near their head and looking toward the conversation partner.

\paragraph{Parameters.}
\begin{center}
\footnotesize
\begin{tabular}{llll}
\toprule
\textbf{Parameter} & \textbf{Type} & \textbf{Default} & \textbf{Description} \\
\midrule
\code{from_actor_name} & str & (required) & Source actor (viewpoint origin) \\
\code{to_actor_name}   & str & (required) & Target actor \\
\code{head_bone_name}  & str & \texttt{head} & Reference head bone \\
\code{forward_offset}   & float & $-20.0$ & Offset along forward axis from source head \\
\code{side_offset}      & float & $50.0$  & Lateral offset from forward axis \\
\bottomrule
\end{tabular}
\end{center}

\paragraph{Geometric computation.}
Using the same forward/right basis as OTS:
\begin{equation}
    \mathbf{p}_{\text{cam}} = \mathbf{b}_F^{(\text{head})}
        + \hat{\mathbf{f}} \cdot d_{\text{fwd}}
        + \hat{\mathbf{r}} \cdot d_{\text{side}},
    \qquad
    \mathbf{t}_{\text{look}} = \mathbf{b}_T^{(\text{head})}
\end{equation}
The negative default of $d_{\text{fwd}} = -20$ places the camera slightly behind the source actor's head, producing a near-POV perspective rather than a literal eye-level shot.

\subsubsection{On-Axis}

The on-axis template places the camera at the midpoint between two characters' heads, creating a direct frontal view of the target subject.

\paragraph{Parameters.}
\begin{center}
\footnotesize
\begin{tabular}{llll}
\toprule
\textbf{Parameter} & \textbf{Type} & \textbf{Default} & \textbf{Description} \\
\midrule
\code{from_actor_name} & str & (required) & Reference actor (midpoint computation) \\
\code{to_actor_name}   & str & (required) & Target actor (camera faces this actor) \\
\code{head_bone_name}  & str & \texttt{head} & Reference head bone \\
\bottomrule
\end{tabular}
\end{center}

\paragraph{Geometric computation.}
\begin{equation}
    \mathbf{p}_{\text{cam}} = \tfrac{1}{2}(\mathbf{b}_F^{(\text{head})} + \mathbf{b}_T^{(\text{head})}),
    \qquad
    \mathbf{t}_{\text{look}} = \mathbf{b}_T^{(\text{head})}
\end{equation}

\subsubsection{Side Profile}

Captures a character from the side, perpendicular to their facing direction.

\paragraph{Parameters.}
\begin{center}
\footnotesize
\begin{tabular}{llll}
\toprule
\textbf{Parameter} & \textbf{Type} & \textbf{Default} & \textbf{Description} \\
\midrule
\code{actor_name}    & str & (required) & Target actor \\
\texttt{side}           & \texttt{left$|$right} & \texttt{left} & Side of the profile \\
\code{side_distance} & float & $300.0$ & Distance from actor \\
\code{bone_name}     & str & \code{spine_03} & Bone for vertical alignment \\
\bottomrule
\end{tabular}
\end{center}

\paragraph{Geometric computation.}
Let $\hat{\mathbf{r}}_A$ be the actor's local right vector (derived from its transform rotation applied to $(-1, 0, 0)$). Then:
\begin{equation}
    \hat{\mathbf{s}} = \begin{cases} -\hat{\mathbf{r}}_A & \text{if side} = \texttt{left} \\ \hat{\mathbf{r}}_A & \text{if side} = \texttt{right} \end{cases}
\end{equation}
\begin{equation}
    \mathbf{p}_{\text{cam}} = \mathbf{b}_A^{(\text{bone})} + \hat{\mathbf{s}} \cdot d_{\text{side}},
    \qquad
    \mathbf{t}_{\text{look}} = \mathbf{b}_A^{(\text{bone})}
\end{equation}

\subsubsection{Establishing Shot}

A wide shot showing multiple characters and their spatial relationship, typically used at scene beginnings.

\paragraph{Parameters.}
\begin{center}
\footnotesize
\begin{tabular}{llll}
\toprule
\textbf{Parameter} & \textbf{Type} & \textbf{Default} & \textbf{Description} \\
\midrule
\code{actor1_name}   & str & (required) & First actor \\
\code{actor2_name}   & str & (required) & Second actor \\
\texttt{side}           & \texttt{left$|$right} & \texttt{right} & Side relative to actor1$\to$actor2 axis \\
\texttt{distance}       & float & $300.0$ & Distance from midpoint \\
\code{height_offset} & float & $150.0$ & Vertical elevation \\
\bottomrule
\end{tabular}
\end{center}

\paragraph{Geometric computation.}
Let $\mathbf{m} = \frac{1}{2}(\mathbf{p}_1 + \mathbf{p}_2)$ be the midpoint and $\hat{\mathbf{d}} = \widehat{\mathbf{p}_2 - \mathbf{p}_1}$ the inter-actor direction. The perpendicular offset vector is:
\begin{equation}
    \hat{\mathbf{r}} = \widehat{\hat{\mathbf{z}} \times \hat{\mathbf{d}}},
    \qquad
    \hat{\mathbf{o}} = \begin{cases} -\hat{\mathbf{r}} & \text{if side} = \texttt{left} \\ \hat{\mathbf{r}} & \text{if side} = \texttt{right} \end{cases}
\end{equation}
\begin{equation}
    \mathbf{p}_{\text{cam}} = \mathbf{m} + \hat{\mathbf{o}} \cdot d + (0, 0, h),
    \qquad
    \mathbf{t}_{\text{look}} = \mathbf{m} + (0, 0, h)
\end{equation}

\subsubsection{Generic Focus}

A flexible single-target template using spherical coordinates relative to the actor's facing direction, suitable for arbitrary framing when the standard templates do not apply.

\paragraph{Parameters.}
\begin{center}
\footnotesize
\begin{tabular}{llll}
\toprule
\textbf{Parameter} & \textbf{Type} & \textbf{Default} & \textbf{Description} \\
\midrule
\code{actor_name} & str   & (required) & Target actor \\
\texttt{distance}    & float & $300.0$ & Camera-to-target distance \\
\texttt{pitch}       & float & $0.0$ & Elevation angle ($>0$: bird's eye, $<0$: worm's eye) \\
\texttt{yaw}         & float & $0.0$ & Horizontal angle rel.\ to actor forward ($0$: front, $\pm 90$: sides, $180$: back)  \\
\code{bone_name}  & str   & \texttt{None} & Optional bone as look-at target \\
\bottomrule
\end{tabular}
\end{center}

\paragraph{Geometric computation.}
Let $\hat{\mathbf{f}}_A$ be the actor's forward vector and define the base direction $\hat{\mathbf{d}}_0 = -\hat{\mathbf{f}}_A$ (facing the actor's front). The camera direction is computed by two successive rotations:
\begin{align}
    \hat{\mathbf{d}}_1 &= \textsc{Rot}(\hat{\mathbf{d}}_0,\; \texttt{yaw},\; \hat{\mathbf{z}}) \\[4pt]
    \hat{\mathbf{r}}_1 &= \widehat{\hat{\mathbf{z}} \times \hat{\mathbf{d}}_1} \\[4pt]
    \hat{\mathbf{d}}_2 &= \textsc{Rot}(\hat{\mathbf{d}}_1,\; -\texttt{pitch},\; \hat{\mathbf{r}}_1)
\end{align}
The camera position is then:
\begin{equation}
    \mathbf{p}_{\text{cam}} = \mathbf{t}_{\text{target}} + \hat{\mathbf{d}}_2 \cdot d
\end{equation}
where $\mathbf{t}_{\text{target}}$ is either $\mathbf{b}_A^{(\text{bone})}$ (if \code{bone_name} is specified) or $\mathbf{p}_A$ (actor root).

\subsection{Movement Templates}
\label{app:movement_templates}

Movement templates animate the camera along a trajectory over a specified duration while maintaining focus on the look-at target computed by the position template.
Both templates take the initial camera pose $(\mathbf{p}_0, \mathbf{R}_0, \mathbf{t}_{\text{look}})$ produced by the position template as input.

\subsubsection{Dolly}

The Dolly template moves the camera toward or away from the subject along the view axis, producing a push-in or pull-out effect.

\paragraph{Parameters.}
\begin{center}
\footnotesize
\begin{tabular}{llll}
\toprule
\textbf{Parameter} & \textbf{Type} & \textbf{Default} & \textbf{Description} \\
\midrule
\texttt{ratio} & float & $0.8$ & Distance scale factor ($< 1$: push in, $> 1$: pull out) \\
\bottomrule
\end{tabular}
\end{center}

\paragraph{Keyframe generation.}
The Dolly template produces exactly two keyframes. Let $\vec{\mathbf{v}} = \mathbf{p}_0 - \mathbf{t}_{\text{look}}$ be the vector from target to initial camera position:
\begin{align}
    \mathbf{p}_{\text{end}} &= \mathbf{t}_{\text{look}} + \vec{\mathbf{v}} \cdot \rho \\[4pt]
    \mathbf{R}_{\text{end}} &= \textsc{LookAt}(\mathbf{p}_{\text{end}},\; \mathbf{t}_{\text{look}})
\end{align}
\begin{center}
\small
\begin{tabular}{cccl}
\toprule
\textbf{Keyframe} & \textbf{Time} & \textbf{Position / Rotation} & \textbf{Interpolation} \\
\midrule
0 & $t_0$          & $\mathbf{p}_0,\; \mathbf{R}_0$                       & \textsc{Linear} \\
1 & $t_0 + T$      & $\mathbf{p}_{\text{end}},\; \mathbf{R}_{\text{end}}$ & \textsc{Constant} \\
\bottomrule
\end{tabular}
\end{center}

The linear interpolation between frame~0 and frame~1 produces a smooth, constant-speed dolly motion. The final \textsc{Constant} key ensures the camera holds its end position.

\subsubsection{Orbit}

The Orbit template rotates the camera around the look-at target along a circular arc on the horizontal plane.

\paragraph{Parameters.}
\begin{center}
\footnotesize
\begin{tabular}{llll}
\toprule
\textbf{Parameter} & \textbf{Type} & \textbf{Default} & \textbf{Description} \\
\midrule
\texttt{angle}     & float & $45.0$  & Total rotation angle (degrees) \\
\texttt{clockwise} & bool  & \texttt{true} & Rotation direction \\
\bottomrule
\end{tabular}
\end{center}

\paragraph{Keyframe generation.}
Let $\vec{\mathbf{r}}_0 = \mathbf{p}_0 - \mathbf{t}_{\text{look}}$ be the initial radius vector and $\theta_{\text{total}} = \alpha$ if clockwise, $-\alpha$ otherwise. The template generates $N = \lfloor T \cdot 30 \rfloor$ intermediate keyframes (at 30\,fps density) to ensure a smooth circular arc:
\begin{align}
    \text{For } i &= 1, \ldots, N{-}1: \notag\\[2pt]
    \theta_i &= \theta_{\text{total}} \cdot \tfrac{i}{N} \\[4pt]
    \vec{\mathbf{r}}_i &= \textsc{Rot}(\vec{\mathbf{r}}_0,\; \theta_i,\; \hat{\mathbf{z}}) \\[4pt]
    \mathbf{p}_i &= \mathbf{t}_{\text{look}} + \vec{\mathbf{r}}_i \\[4pt]
    \mathbf{R}_i &= \textsc{LookAt}(\mathbf{p}_i,\; \mathbf{t}_{\text{look}})
\end{align}
Each intermediate keyframe uses \textsc{Linear} interpolation; the final keyframe at $t_0 + T$ uses \textsc{Constant} to hold the end pose. The constant angular sampling ensures uniform angular velocity throughout the orbit.


\section{CutsceneBench: Example Test Case}
\label{app:test_scenarios}

\definecolor{shotcolor}{RGB}{0, 80, 160}
\definecolor{dialogcolor}{RGB}{140, 30, 30}
\definecolor{actioncolor}{RGB}{40, 110, 50}
\definecolor{phasecolor}{RGB}{50, 50, 130}

We present a complete S2-tier test case (\textit{Checkpoint Negotiation}) to illustrate the structure of CutsceneBench data.
Each test case comprises three components:
\begin{enumerate}[nosep, leftmargin=*]
  \item A \textbf{storyboard prompt}---the sole input provided to the cutscene agent;
  \item A \textbf{ground-truth tool-call trajectory}---the reference sequence of MCP tool invocations used for Layer~1 evaluation;
  \item A \textbf{ground-truth Level Sequence snapshot}---the expected final timeline state used for Layer~2 evaluation.
\end{enumerate}
Only the storyboard~(1) is visible to the agent during generation; components~(2) and~(3) serve exclusively as evaluation references.

\subsection{Storyboard Prompt}
\label{sec:app_storyboard}

The storyboard below is provided verbatim as the initial user message.
It specifies participating characters, spatial staging, shot-by-shot camera direction, dialogue, and acting cues.

\begin{tcolorbox}[
  title={\textbf{S2\_001 --- Checkpoint Negotiation}},
  colback=accentlight, colframe=accentmid,
  fonttitle=\small\bfseries, arc=3pt,
  left=6pt, right=6pt, top=4pt, bottom=4pt,
]
\small
\textcolor{black!50}{\textsf{Game}} \enspace \textit{The Lost Courier} \hfill
\textcolor{black!50}{\textsf{ID}} \enspace CUT\_02\_007 \hfill
\textcolor{black!50}{\textsf{Duration}} \enspace $\sim$30\,s \hfill
\textcolor{black!50}{\textsf{Tier}} \enspace S2 \\[2pt]
\textcolor{black!50}{\textsf{Trigger}} \enspace Player's first attempt to pass the North Gate checkpoint. \\[6pt]
\textcolor{black!50}{\textsf{Characters}}
\begin{itemize}[nosep,leftmargin=12pt]
  \item \textbf{MIRA} --- standing, facing REX
  \item \textbf{REX} --- standing, facing MIRA, holding a registry clipboard
\end{itemize}
\textcolor{black!50}{\textsf{Location}} \enspace Two sides of a checkpoint iron fence, daytime.
\tcbline
\textcolor{shotcolor}{\textbf{Shot 01}} \enspace {\footnotesize\textsf{Medium two-shot\,|\,static\,|\,$\sim$4\,s}} \\
\textcolor{actioncolor}{Staging: MIRA left, REX right, fence centered.} \\
\textcolor{actioncolor}{Action: REX looks down at clipboard; MIRA stands waiting.} \\[5pt]
\textcolor{shotcolor}{\textbf{Shot 02}} \enspace {\footnotesize\textsf{OTS from REX onto MIRA\,|\,static\,|\,$\sim$7\,s}} \\
\textbf{REX} \textit{\small(without looking up):} \textcolor{dialogcolor}{``Pass and destination.''} \\
\textbf{MIRA} \textit{\small(takes a slip from coat, passes through fence):} \textcolor{dialogcolor}{``East district. Courier.''} \\
\textcolor{actioncolor}{Action: MIRA extends arm to hand over the pass; REX takes it, looks down.} \\[5pt]
\textcolor{shotcolor}{\textbf{Shot 03}} \enspace {\footnotesize\textsf{OTS from MIRA onto REX\,|\,static\,|\,$\sim$6\,s}} \\
\textbf{REX} \textit{\small(looks up):} \textcolor{dialogcolor}{``Courier's guild?''} \qquad
\textbf{MIRA}: \textcolor{dialogcolor}{``Freelance.''} \\[5pt]
\textcolor{shotcolor}{\textbf{Shot 04}} \enspace {\footnotesize\textsf{Medium two-shot\,|\,static\,|\,$\sim$8\,s}} \\
\textbf{REX} \textit{\small(looking down, hands pass back):} \textcolor{dialogcolor}{``Go ahead. East district curfew is at six---don't be late.''} \\
\textbf{MIRA} \textit{\small(takes pass, nods):} \textcolor{dialogcolor}{``Understood.''} \\
\textcolor{actioncolor}{Action: REX hands back the pass and steps aside; MIRA walks forward.} \\[5pt]
\textcolor{shotcolor}{\textbf{Shot 05}} \enspace {\footnotesize\textsf{Medium shot\,|\,static\,|\,$\sim$5\,s}} \\
\textcolor{actioncolor}{REX alone by the fence; looks down at clipboard again.}
\end{tcolorbox}

\subsection{Ground-Truth Tool-Call Trajectory}
\label{sec:app_trajectory}

The reference trajectory records the expected sequence of MCP tool invocations.
This case requires approximately 45 calls organised across five execution phases; calls within the same phase may be issued in parallel.

\begin{tcolorbox}[
  title={\textbf{Reference Trajectory --- S2\_001}},
  colback=gray!3, colframe=gray!50!black,
  fonttitle=\small\bfseries, arc=3pt,
  left=6pt, right=6pt, top=3pt, bottom=3pt,
  breakable,
]
\small

\textcolor{phasecolor}{\textsf{\textbf{Phase 1: Asset Discovery}}} \hfill {\footnotesize\textit{3 calls}}\\[2pt]
\quad \texttt{get\_available\_characters()} \\
\quad \texttt{get\_available\_tone()} \\
\quad \texttt{get\_available\_camera\_templates()} \\[6pt]
\textcolor{phasecolor}{\textsf{\textbf{Phase 2: Character Setup}}} \hfill {\footnotesize\textit{3 calls}}\\[2pt]
\quad \texttt{add\_character}(\textit{name}=\texttt{"MIRA"},\enspace \textit{id}=\texttt{"char\_001"},\enspace \textit{loc}=[-60,0,0]) \\
\quad \texttt{add\_character}(\textit{name}=\texttt{"REX"},\enspace \textit{id}=\texttt{"char\_002"},\enspace \textit{loc}=[60,0,0]) \\
\quad \texttt{orient\_character\_to\_center}(\textit{names}=[\texttt{"MIRA"}, \texttt{"REX"}]) \\[6pt]
\textcolor{phasecolor}{\textsf{\textbf{Phase 3: Speech \& Facial Animation}}} \hfill {\footnotesize\textit{12 calls}}\\[2pt]
\quad \texttt{tts\_function}(\textit{id}=\texttt{"rex\_line1\_audio"},\enspace \textit{text}=\texttt{"..."},\enspace \textit{gender}=\texttt{"male"},\enspace \textit{tone}=\texttt{"male\_normal\_1.mp3"}) \\
\quad \quad \textit{... ($\times$6 total, one per dialogue line)} \\[2pt]
\quad \texttt{audio\_to\_face\_expression}(\textit{id}=\texttt{"rex\_line1\_face"},\enspace \textit{audio\_id}=\texttt{"rex\_line1\_audio"}) \\
\quad \quad \textit{... ($\times$6 total, one per audio asset)} \\[6pt]
\textcolor{phasecolor}{\textsf{\textbf{Phase 4: Track Assembly}}} \hfill {\footnotesize\textit{$\sim$27 calls}}\\[2pt]
\quad \texttt{add\_character\_audio}(\textit{char}=\texttt{"REX"},\enspace \textit{id}=\texttt{"rex\_line1\_audio"},\enspace \textit{start}=4.5,\enspace \textit{end}=6.33) \\
\quad \quad \textit{... ($\times$6 total, attaching audio sections to character tracks)} \\[2pt]
\quad \texttt{add\_character\_facial\_animation}(\textit{char}=\texttt{"REX"},\enspace \textit{id}=\texttt{"rex\_line1\_face"},\enspace \textit{start}=4.5) \\
\quad \quad \textit{... ($\times$6 total)} \\[2pt]
\quad \texttt{get\_available\_animations}(\textit{gender}=\texttt{"male"}) \\
\quad \texttt{add\_character\_animation}(\textit{char}=\texttt{"REX"},\enspace \textit{id}=\texttt{"A\_M\_Closeup\_Guard"},\enspace \textit{start}=0.0) \\
\quad \quad \textit{... ($\times$15 total, continuous body-animation coverage for both characters)} \\[6pt]
\textcolor{phasecolor}{\textsf{\textbf{Phase 5: Camera Setup}}} \hfill {\footnotesize\textit{15 calls}}\\[2pt]
\quad \texttt{add\_camera}(\textit{name}=\texttt{"Cam\_OTS\_Rex"}) \quad\textit{... ($\times$5 cameras)} \\[2pt]
\quad \texttt{apply\_camera\_template}(\textit{cam}=\texttt{"Cam\_OTS\_Rex"},\enspace \textit{template}=\texttt{"OTS"},\\
\quad \quad \textit{args}=\{\textit{from}: \texttt{"REX"}, \textit{to}: \texttt{"MIRA"}\},\enspace \textit{start}=4.0,\enspace \textit{dur}=7.0)\\
\quad \quad \textit{... ($\times$5 total)} \\[2pt]
\quad \texttt{set\_active\_camera}(\textit{cam}=\texttt{"Cam\_OTS\_Rex"},\enspace \textit{start}=4.0,\enspace \textit{end}=11.0) \\
\quad \quad \textit{... ($\times$5 total, continuous camera-cut coverage)}

\end{tcolorbox}

\subsection{Ground-Truth Level Sequence Snapshot}
\label{sec:app_snapshot}

The Level Sequence snapshot captures the expected final state of the Unreal Engine timeline after all tool calls complete.
Figure~\ref{fig:level_seq_structure} shows the hierarchical track layout for this case.

\begin{tcolorbox}[
  title={\textbf{Level Sequence Snapshot} {\normalfont(simplified)}},
  colback=black!2, colframe=black!50,
  fonttitle=\small\bfseries, arc=3pt,
  left=6pt, right=6pt, top=4pt, bottom=4pt,
]
\small

\textbf{Camera Bindings} \enspace\textit{(5 CineCameraActors)} \\[2pt]
\quad Cam\_Establishing,\enspace Cam\_OTS\_Rex,\enspace Cam\_OTS\_Mira,\enspace Cam\_Establishing2,\enspace Cam\_SideProfile\_Rex \\[8pt]
\textbf{Character: REX}
\begin{itemize}[nosep, leftmargin=15pt, label={\small$\triangleright$}]
  \item \textit{Audio Track} --- 3 sections:\enspace [4.5--6.3\,s]\enspace [11.5--12.7\,s]\enspace [17.5--20.4\,s]
  \item \textit{Facial Animation Track} --- 3 sections (temporally aligned with audio)
  \item \textit{Body Animation Track} --- 8 sections, continuous coverage [0--30\,s]
\end{itemize}
\vspace{4pt}
\textbf{Character: MIRA}
\begin{itemize}[nosep, leftmargin=15pt, label={\small$\triangleright$}]
  \item \textit{Audio Track} --- 3 sections:\enspace [7.0--8.8\,s]\enspace [13.5--15.1\,s]\enspace [21.5--23.9\,s]
  \item \textit{Facial Animation Track} --- 3 sections (temporally aligned with audio)
  \item \textit{Body Animation Track} --- 7 sections, continuous coverage [0--30\,s]
\end{itemize}
\vspace{6pt}
\textbf{Camera Cut Track} \enspace\textit{(5 sections, continuous)} \\[2pt]
\quad Establishing\,[0--4\,s] \;$\rightarrow$\; OTS\_Rex\,[4--11\,s] \;$\rightarrow$\; OTS\_Mira\,[11--17\,s] \\
\quad $\rightarrow$\; Establishing2\,[17--25\,s] \;$\rightarrow$\; SideProfile\_Rex\,[25--30\,s]

\end{tcolorbox}
\captionof{figure}{Hierarchical structure of the Level Sequence produced by S2\_001.
  Two characters each carry audio, facial animation, and body animation tracks;
  five cameras provide continuous coverage via the camera-cut track.}
\label{fig:level_seq_structure}


\section{Evaluation Details}
\label{app:eval_details}

\subsection{Layer 3 Evaluation Prompt}
\label{app:l3_prompt}

The following prompt is provided to the multimodal judge model (Gemini 3.1 Pro) alongside the rendered cutscene video and the original storyboard script. The judge evaluates four dimensions independently, producing chain-of-thought reasoning and a discrete score (0--25) for each.

\begin{tcolorbox}[
  title={\textbf{L3 Judge Prompt}},
  colback=gray!2, colframe=gray!60!black,
  fonttitle=\small\bfseries, arc=3pt,
  left=6pt, right=6pt, top=4pt, bottom=4pt,
  breakable
]
\small
You will be given a rendered cutscene video from a video game, along with the original storyboard script that was used as input to an AI agent that generated this cutscene.

Your task is to evaluate the quality of the generated cutscene across four independent dimensions, each scored from 0 to 25 (total 100 points).

\tcbline
\textbf{EVALUATION DIMENSIONS AND RUBRICS} \\[4pt]
\textcolor{accentdark}{\textbf{1.\ Script Fidelity (SF)}} \\
\textit{Focus:} Whether dialogue content and character actions specified in the storyboard are accurately reproduced.

\begin{tabular}{@{}rl@{}}
  0--5:  & No recognizable connection to the script; characters or dialogue entirely wrong. \\
  6--10: & Major dialogue lines missing or assigned to wrong characters; key actions absent. \\
 11--15: & Most dialogue present but with notable omissions; some actions missing. \\
 16--20: & All dialogue present and correctly assigned; minor action inaccuracies. \\
 21--25: & Dialogue and actions faithfully and completely match the script. \\
\end{tabular}

\vspace{4pt}
\textcolor{accentdark}{\textbf{2.\ Character Consistency (ChC)}} \\
\textit{Focus:} Whether characters maintain stable identities, spatial positions, and coherent behavior throughout.

\begin{tabular}{@{}rl@{}}
  0--5:  & Characters constantly glitch, teleport, or are unrecognizable. \\
  6--10: & Significant issues---characters swap positions, face wrong directions frequently. \\
 11--15: & Generally stable but with noticeable position jumps or animation stiffness. \\
 16--20: & Consistent presentation with only minor issues (slight clipping, brief stiffness). \\
 21--25: & Fully consistent; characters behave naturally throughout. \\
\end{tabular}

\vspace{4pt}
\textcolor{accentdark}{\textbf{3.\ Cinematographic Quality (CQ)}} \\
\textit{Focus:} Whether camera shot types, framing, and cutting patterns serve the narrative effectively.

\begin{tabular}{@{}rl@{}}
  0--5:  & Camera is broken, static, or completely fails to frame the action. \\
  6--10: & Poor shot selection; characters frequently out of frame or awkwardly framed. \\
 11--15: & Adequate framing but shot types don't match the script; some jarring cuts. \\
 16--20: & Good shot selection matching the script; characters well-framed with minor issues. \\
 21--25: & Professional-quality cinematography; shots and cuts enhance the narrative. \\
\end{tabular}

\vspace{4pt}
\textcolor{accentdark}{\textbf{4.\ Temporal Coherence (TmpCoh)}} \\
\textit{Focus:} Whether timing and synchronization create a natural viewing experience.

\begin{tabular}{@{}rl@{}}
  0--5:  & Completely broken timing; audio and animation fully desynchronized. \\
  6--10: & Significant issues---dialogue stutters/repeats, long dead air, animations freeze. \\
 11--15: & Noticeable timing problems but generally followable; some sync drift. \\
 16--20: & Smooth pacing with minor timing imperfections. \\
 21--25: & Seamless flow; all elements perfectly synchronized and naturally paced. \\
\end{tabular}

\tcbline
\textbf{EVALUATION STEPS} \\[2pt]
For \textbf{each} dimension independently: \\
\quad Step 1: Watch the video carefully, focusing ONLY on aspects relevant to that dimension. \\
\quad Step 2: Compare observations against the storyboard and the rubric above. \\
\quad Step 3: Write chain-of-thought reasoning (2--4 sentences). \\
\quad Step 4: Assign an integer score from 0 to 25.

\tcbline
\textbf{OUTPUT FORMAT} --- Respond with ONLY a valid JSON object:

\begin{verbatim}
{"script_fidelity": {"reasoning": "...", "score": 20},
 "character_consistency": {"reasoning": "...", "score": 22},
 "cinematographic_quality": {"reasoning": "...", "score": 18},
 "temporal_coherence": {"reasoning": "...", "score": 15}}
\end{verbatim}
\end{tcolorbox}

\subsection{Full Evaluation Results}
\label{app:full_results}

This section provides comprehensive per-scenario breakdowns across all three evaluation layers. Note that Qwen 2.5-72B is excluded from all tables in this section: as detailed in Section~\ref{sec:eval_results}, its substantially lower L1--L2 scores (e.g., CC 56.6\%, TC 50.9\%) indicate a qualitative performance gap on the cutscene generation task, and the resulting Level Sequences lack sufficient structural integrity for meaningful L3 evaluation.

\subsubsection{Combined Overall Summary}

Table~\ref{tab:combined_overall} consolidates results across all evaluation layers. L1 and L2 metrics are averaged over 65 test scenarios per model (reported in \%); L3 metrics are averaged over 25 sampled videos per model (scored 0--25 per dimension, 0--100 total).

\begin{table}[h]
\centering
\small
\begin{tabular}{@{}l ccccc ccc c@{}}
\toprule
& \multicolumn{5}{c}{\textbf{Layer 1 (\%)}} & \multicolumn{3}{c}{\textbf{Layer 2 (\%)}} & \textbf{L3} \\
\cmidrule(lr){2-6} \cmidrule(lr){7-9} \cmidrule(lr){10-10}
\textbf{Model} & \textbf{TSA} & \textbf{PV} & \textbf{CC} & \textbf{CE} & \textbf{DC} & \textbf{TC} & \textbf{CamC} & \textbf{TempC} & \textbf{Total} \\
\midrule
Claude Opus 4.6      & 100.0 & 100.0 & 100.0 &  97.5 & 100.0 & 100.0 &  96.4 &  99.5 &  50.2 \\
Claude Sonnet 4.6    & 100.0 &  99.9 &  98.4 &  97.4 & 100.0 &  99.6 &  89.5 &  98.6 &  41.7 \\
GPT-5.4              & 100.0 &  96.6 &  95.7 &  97.4 &  98.5 &  96.0 &  93.5 &  98.0 &  42.4 \\
Qwen 3.5 Plus        &  99.9 &  97.1 &  94.5 &  99.7 &  99.5 &  97.9 &  89.3 &  96.3 &  30.0 \\
Kimi K2.5            &  99.3 &  97.4 &  91.8 &  98.6 &  98.7 &  91.0 &  73.9 &  89.2 &  30.7 \\
GLM-5                &  99.7 &  98.1 &  93.1 &  99.2 &  99.2 &  92.4 &  77.3 &  95.8 &  28.9 \\
MiniMax M2.5         &  99.5 &  91.6 &  90.9 &  98.7 &  99.2 &  94.8 &  74.8 &  85.3 &  25.8 \\
\bottomrule
\end{tabular}
\caption{Combined evaluation results across all layers. L1--L2 values in \%; L3 Total scored 0--100. Models are ordered by L3 Total.}
\label{tab:combined_overall}
\end{table}

\subsubsection{Per-Scenario Complexity Scaling}

Tables~\ref{tab:scaling_cc}--\ref{tab:scaling_l3} report per-scenario breakdowns for the three most discriminative metrics: Call Completeness (L1), Camera Coverage (L2), and L3 Total.

\begin{table}[h]
\centering
\small
\begin{tabular}{@{}lcccccc@{}}
\toprule
\textbf{Model} & \textbf{S1} & \textbf{S2} & \textbf{S3} & \textbf{S4} & \textbf{S5} & $\boldsymbol{\Delta}$ \\
\midrule
Claude Opus 4.6      & 100.0 & 100.0 & 100.0 & 100.0 & 100.0 &  0.0 \\
Claude Sonnet 4.6    & 100.0 &  99.5 &  94.9 &  99.2 &  92.3 & $-$7.7 \\
GPT-5.4              &  95.9 &  99.5 &  92.5 &  94.6 &  87.5 & $-$8.4 \\
Qwen 3.5 Plus        &  94.7 &  97.4 &  95.4 &  92.4 &  84.9 & $-$9.8 \\
Kimi K2.5            &  89.8 &  97.8 &  88.8 &  94.3 &  77.1 & $-$12.7 \\
GLM-5                &  95.6 &  94.5 &  88.8 &  94.7 &  83.1 & $-$12.5 \\
MiniMax M2.5         &  92.6 &  92.4 &  88.3 &  86.0 &  87.2 & $-$5.4 \\
\bottomrule
\end{tabular}
\caption{Call Completeness (\%) by scenario tier. $\Delta$ = S5 $-$ S1. S1: single-character (20 cases), S2: two-person dialogue (20), S3: emotional scene (10), S4: three-person (10), S5: complex multi-turn (5).}
\label{tab:scaling_cc}
\end{table}

\begin{table}[h]
\centering
\small
\begin{tabular}{@{}lcccccc@{}}
\toprule
\textbf{Model} & \textbf{S1} & \textbf{S2} & \textbf{S3} & \textbf{S4} & \textbf{S5} & $\boldsymbol{\Delta}$ \\
\midrule
Claude Opus 4.6      &  94.7 &  96.7 &  98.5 &  97.2 &  96.7 &   2.0 \\
Claude Sonnet 4.6    &  81.7 &  90.6 &  95.1 &  95.5 &  93.3 &  11.6 \\
GPT-5.4              &  93.4 &  91.9 &  95.2 &  94.0 &  96.3 &   2.9 \\
Qwen 3.5 Plus        &  90.0 &  88.2 &  88.9 &  92.6 &  85.1 & $-$4.9 \\
Kimi K2.5            &  75.5 &  74.9 &  82.8 &  67.1 &  59.4 & $-$16.1 \\
GLM-5                &  76.1 &  75.5 &  84.7 &  67.2 &  95.3 &  19.2 \\
MiniMax M2.5         &  77.0 &  73.5 &  82.4 &  51.6 &  96.5 &  19.5 \\
\bottomrule
\end{tabular}
\caption{Camera Coverage (\%) by scenario tier. $\Delta$ = S5 $-$ S1. Note: S5 sample sizes for MiniMax (2 cases) and S4 for MiniMax (6 cases) are smaller due to missing sessions, so those values have higher variance.}
\label{tab:scaling_camc}
\end{table}

\begin{table}[h]
\centering
\small
\begin{tabular}{@{}lcccccc@{}}
\toprule
\textbf{Model} & \textbf{S1} & \textbf{S2} & \textbf{S3} & \textbf{S4} & \textbf{S5} & $\boldsymbol{\Delta}$ \\
\midrule
Claude Opus 4.6      &  61.4 &  51.0 &  61.4 &  36.6 &  40.6 & $-$20.8 \\
Claude Sonnet 4.6    &  40.8 &  49.8 &  45.8 &  37.8 &  34.4 & $-$6.4 \\
GPT-5.4              &  45.2 &  42.2 &  62.4 &  20.0 &  37.6 & $-$7.6 \\
Qwen 3.5 Plus        &  24.8 &  40.0 &  35.6 &  25.8 &  23.6 & $-$1.2 \\
Kimi K2.5            &  19.2 &  43.0 &  36.0 &  28.0 &  27.2 &   8.0 \\
GLM-5                &  18.8 &  36.6 &  52.0 &  19.4 &  17.6 & $-$1.2 \\
MiniMax M2.5         &  20.4 &  34.6 &  34.0 &  16.8 &  23.0 &   2.6 \\
\bottomrule
\end{tabular}
\caption{Layer 3 total score (0--100) by scenario tier, averaged over 5 videos per tier. $\Delta$ = S5 $-$ S1.}
\label{tab:scaling_l3}
\end{table}

\subsubsection{Full Per-Scenario L1 Breakdown}

Table~\ref{tab:full_l1_per_scenario} provides the complete L1 per-scenario breakdown for all five metrics.

\begin{table}[h]
\centering
\setlength{\tabcolsep}{3.5pt}
\footnotesize
\begin{tabular}{@{}l ccccc ccccc ccccc@{}}
\toprule
& \multicolumn{5}{c}{\textbf{S1 (20 cases)}} & \multicolumn{5}{c}{\textbf{S2 (20 cases)}} & \multicolumn{5}{c}{\textbf{S3 (10 cases)}} \\
\cmidrule(lr){2-6} \cmidrule(lr){7-11} \cmidrule(lr){12-16}
\textbf{Model} & TSA & PV & CC & CE & DC & TSA & PV & CC & CE & DC & TSA & PV & CC & CE & DC \\
\midrule
Opus 4.6 & 100 & 100 & 100 & 98.3 & 100 & 100 & 100 & 100 & 97.0 & 100 & 100 & 100 & 100 & 97.0 & 100 \\
Sonnet 4.6 & 100 & 99.8 & 100 & 98.0 & 100 & 100 & 99.8 & 99.5 & 96.9 & 100 & 100 & 100 & 94.9 & 96.9 & 100 \\
GPT-5.4 & 100 & 94.8 & 95.9 & 97.7 & 99.0 & 100 & 96.5 & 99.5 & 97.5 & 97.0 & 100 & 97.6 & 92.5 & 96.8 & 98.8 \\
Qwen 3.5+ & 99.8 & 96.3 & 94.7 & 100 & 99.0 & 100 & 97.1 & 97.4 & 99.8 & 99.8 & 100 & 97.5 & 95.4 & 100 & 99.2 \\
Kimi K2.5 & 98.4 & 95.8 & 89.8 & 99.2 & 97.2 & 99.7 & 97.3 & 97.8 & 98.4 & 99.3 & 99.7 & 98.5 & 88.8 & 98.3 & 99.6 \\
GLM-5   & 99.2 & 99.1 & 95.6 & 99.2 & 99.0 & 99.9 & 98.3 & 94.5 & 100 & 99.2 & 99.8 & 98.7 & 88.8 & 98.9 & 99.4 \\
MiniMax M2.5 & 98.8 & 91.4 & 92.6 & 98.4 & 99.0 & 99.8 & 89.4 & 92.4 & 99.3 & 99.3 & 100 & 91.4 & 88.3 & 98.0 & 99.3 \\
\bottomrule
\end{tabular}

\vspace{6pt}

\begin{tabular}{@{}l ccccc ccccc@{}}
\toprule
& \multicolumn{5}{c}{\textbf{S4 (10 cases)}} & \multicolumn{5}{c}{\textbf{S5 (5 cases)}} \\
\cmidrule(lr){2-6} \cmidrule(lr){7-11}
\textbf{Model} & TSA & PV & CC & CE & DC & TSA & PV & CC & CE & DC \\
\midrule
Opus 4.6 & 100 & 100 & 100 & 97.7 & 100 & 100 & 100 & 100 & 97.5 & 100 \\
Sonnet 4.6 & 100 & 100 & 99.2 & 97.6 & 100 & 100 & 100 & 92.3 & 97.5 & 100 \\
GPT-5.4 & 100 & 97.8 & 94.6 & 97.2 & 99.5 & 100 & 99.3 & 87.5 & 97.5 & 100 \\
Qwen 3.5+ & 100 & 99.1 & 92.4 & 99.2 & 99.8 & 100 & 95.3 & 84.9 & 98.4 & 100 \\
Kimi K2.5 & 99.8 & 99.0 & 94.3 & 97.7 & 99.0 & 100 & 99.0 & 77.1 & 98.5 & 100 \\
GLM-5   & 100 & 96.9 & 94.7 & 97.7 & 99.5 & 100 & 94.4 & 83.1 & 99.0 & 98.9 \\
MiniMax M2.5 & 100 & 96.5 & 86.0 & 98.8 & 99.1 & 100 & 99.3 & 87.2 & 100 & 100 \\
\bottomrule
\end{tabular}
\caption{Full L1 per-scenario results (\%). MiniMax M2.5 has reduced case counts in S4 (6) and S5 (2) due to missing sessions.}
\label{tab:full_l1_per_scenario}
\end{table}

\subsubsection{Full Per-Scenario L2 Breakdown}

\begin{table}[H]
\centering
\resizebox{\textwidth}{!}{%
\begin{tabular}{@{}l ccc ccc ccc ccc ccc@{}}
\toprule
& \multicolumn{3}{c}{\textbf{S1}} & \multicolumn{3}{c}{\textbf{S2}} & \multicolumn{3}{c}{\textbf{S3}} & \multicolumn{3}{c}{\textbf{S4}} & \multicolumn{3}{c}{\textbf{S5}} \\
\cmidrule(lr){2-4} \cmidrule(lr){5-7} \cmidrule(lr){8-10} \cmidrule(lr){11-13} \cmidrule(lr){14-16}
\textbf{Model} & TC & CamC & TmpC & TC & CamC & TmpC & TC & CamC & TmpC & TC & CamC & TmpC & TC & CamC & TmpC \\
\midrule
Opus 4.6 & 100 & 94.7 & 99.6 & 100 & 96.7 & 99.8 & 100 & 98.5 & 97.8 & 100 & 97.2 & 100 & 100 & 96.7 & 100 \\
Sonnet 4.6 & 100 & 81.7 & 100 & 100 & 90.6 & 97.7 & 97.5 & 95.1 & 98.7 & 100 & 95.5 & 97.6 & 100 & 93.3 & 98.2 \\
GPT-5.4 & 95.8 & 93.4 & 97.5 & 100 & 91.9 & 97.2 & 88.9 & 95.2 & 99.3 & 93.7 & 94.0 & 99.5 & 100 & 96.3 & 98.0 \\
Qwen 3.5+ & 100 & 90.0 & 97.0 & 100 & 88.2 & 94.8 & 94.4 & 88.9 & 98.6 & 94.6 & 92.6 & 96.0 & 95.0 & 85.1 & 96.1 \\
Kimi K2.5 & 100 & 75.5 & 85.8 & 96.7 & 74.9 & 91.2 & 69.3 & 82.8 & 89.6 & 90.7 & 67.1 & 90.0 & 76.7 & 59.4 & 92.2 \\
GLM-5   & 100 & 76.1 & 95.2 & 95.0 & 75.5 & 97.3 & 75.7 & 84.7 & 92.7 & 97.1 & 67.2 & 96.8 & 75.4 & 95.3 & 96.8 \\
MiniMax M2.5 & 97.5 & 77.0 & 81.8 & 98.3 & 73.5 & 87.5 & 89.8 & 82.4 & 90.6 & 90.2 & 51.6 & 82.1 & 75.0 & 96.5 & 84.1 \\
\bottomrule
\end{tabular}}
\caption{Full L2 per-scenario results (\%). TmpC = Temporal Consistency.}
\label{tab:full_l2_per_scenario}
\end{table}

\subsubsection{Full Per-Scenario L3 Breakdown}

\begin{table}[H]
\centering
\small
\begin{tabular}{@{}l ccccc ccccc@{}}
\toprule
& \multicolumn{5}{c}{\textbf{S1}} & \multicolumn{5}{c}{\textbf{S2}} \\
\cmidrule(lr){2-6} \cmidrule(lr){7-11}
\textbf{Model} & SF & ChC & CQ & TmpCoh & Tot. & SF & ChC & CQ & TmpCoh & Tot. \\
\midrule
Opus 4.6 & 11.2 & 19.2 & 14.0 & 17.0 & 61.4 & 11.6 & 12.8 & 17.2 &  9.4 & 51.0 \\
Sonnet 4.6 & 11.0 & 10.4 &  7.6 & 11.8 & 40.8 & 11.2 & 14.8 & 15.4 &  8.4 & 49.8 \\
GPT-5.4 & 10.0 & 15.0 &  9.6 & 10.6 & 45.2 &  9.8 & 13.0 & 12.2 &  7.2 & 42.2 \\
Qwen 3.5+ &  7.4 &  7.4 &  2.0 &  8.0 & 24.8 &  8.6 & 12.8 & 11.8 &  6.8 & 40.0 \\
Kimi K2.5 &  6.2 &  6.2 &  1.2 &  5.6 & 19.2 & 10.8 & 12.8 & 10.6 &  8.8 & 43.0 \\
GLM-5   &  6.4 &  3.6 &  1.6 &  7.2 & 18.8 &  9.6 & 11.2 &  8.6 &  7.2 & 36.6 \\
MiniMax M2.5 &  6.6 &  4.6 &  2.0 &  7.2 & 20.4 &  9.6 &  8.2 &  9.2 &  7.6 & 34.6 \\
\bottomrule
\end{tabular}

\vspace{6pt}

\resizebox{\textwidth}{!}{%
\begin{tabular}{@{}l ccccc ccccc ccccc@{}}
\toprule
& \multicolumn{5}{c}{\textbf{S3}} & \multicolumn{5}{c}{\textbf{S4}} & \multicolumn{5}{c}{\textbf{S5}} \\
\cmidrule(lr){2-6} \cmidrule(lr){7-11} \cmidrule(lr){12-16}
\textbf{Model} & SF & ChC & CQ & TC & Tot. & SF & ChC & CQ & TC & Tot. & SF & ChC & CQ & TC & Tot. \\
\midrule
Opus 4.6 & 13.0 & 16.8 & 16.0 & 15.6 & 61.4 &  8.4 & 11.0 &  9.6 &  7.6 & 36.6 & 10.0 & 10.6 &  9.2 & 10.8 & 40.6 \\
Sonnet 4.6 & 10.8 & 12.6 & 10.8 & 11.6 & 45.8 &  9.8 &  9.8 &  8.8 &  9.4 & 37.8 &  9.4 & 10.8 &  6.6 &  7.6 & 34.4 \\
GPT-5.4 & 12.8 & 18.2 & 15.8 & 15.6 & 62.4 &  6.5 &  3.8 &  3.8 &  6.0 & 20.0 & 11.0 &  9.6 &  7.4 &  9.6 & 37.6 \\
Qwen 3.5+ &  7.2 & 14.2 &  5.6 &  8.6 & 35.6 &  7.2 &  7.0 &  5.2 &  6.4 & 25.8 &  6.6 &  7.0 &  4.0 &  6.0 & 23.6 \\
Kimi K2.5 &  8.8 & 12.0 &  7.2 &  8.0 & 36.0 &  8.4 &  9.0 &  4.4 &  6.2 & 28.0 &  8.8 &  7.8 &  3.8 &  6.8 & 27.2 \\
GLM-5   & 11.4 & 17.8 & 11.6 & 11.2 & 52.0 &  6.6 &  4.0 &  3.6 &  5.2 & 19.4 &  6.0 &  5.2 &  3.0 &  3.4 & 17.6 \\
MiniMax M2.5 &  7.6 & 13.8 &  5.0 &  7.6 & 34.0 &  5.8 &  4.6 &  3.2 &  3.2 & 16.8 &  8.0 &  7.0 &  2.6 &  5.4 & 23.0 \\
\bottomrule
\end{tabular}}
\caption{Full L3 per-scenario per-dimension results (scored 0--25; Tot.\ = total 0--100). Averaged over 5 videos per tier per model. TC = TmpCoh (abbreviated for table width).}
\label{tab:full_l3_per_scenario}
\end{table}

\subsubsection{Complexity Scaling Analysis}

\textbf{L1 Call Completeness} degrades monotonically with scenario complexity for all models except Claude Opus 4.6, which maintains perfect coverage across all tiers (Table~\ref{tab:scaling_cc}). The degradation is most pronounced for Kimi K2.5 ($\Delta = -12.7$ pp) and GLM-5 ($\Delta = -12.5$ pp), both falling below 84\% on S5, indicating that context management breaks down for scenes requiring 60+ sequential tool calls. TSA and DC remain nearly perfect across all tiers for all models, confirming that tool selection and dependency awareness are complexity-invariant; the primary bottleneck is operational recall (CC).

\textbf{L2 Camera Coverage} shows a more complex pattern (Table~\ref{tab:scaling_camc}). For top-tier models (Opus, GPT-5.4), CamC remains stable around 93--98\% across all tiers. However, lower-tier models exhibit severe degradation on multi-character scenarios: Kimi K2.5 drops from 75.5\% (S1) to 59.4\% (S5), and MiniMax M2.5 reaches a minimum of 51.6\% on S4, indicating that nearly half the scene has no active camera assigned. Interestingly, the S3 tier (two-character emotional scenes) yields better CamC than S1/S2 for several models, suggesting that richer narrative context helps the agent plan more comprehensive camera coverage.

\textbf{L3 Cinematic Quality} exhibits a non-monotonic relationship with scenario complexity (Table~\ref{tab:scaling_l3}). Several models peak on S2 or S3 rather than S1---e.g., GPT-5.4 achieves 62.4 on S3 but only 45.2 on S1, and Kimi K2.5 scores 43.0 on S2 but only 19.2 on S1. This suggests that cinematic quality depends not only on task complexity but also on \emph{narrative richness}: emotionally expressive dialogue scenes provide more opportunities for meaningful camera work and character performance than simple monologues. Conversely, S4 (three-character scenes) proves the most challenging tier across all models, with GPT-5.4 dropping sharply to 20.0 and MiniMax M2.5 to 16.8.




\end{document}